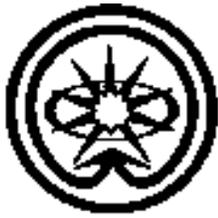

Государственный научный центр РФ
**Институт теоретической и экспериментальной физики**

# ПРОГРАММА

## ИССЛЕДОВАНИЙ НА АДРОННЫХ ПУЧКАХ УСКОРИТЕЛЯ У-10 ИТЭФ

Сборник предложений экспериментов
под редакцией В.В. Владимирского,
В.П. Канавца и В.Т. Смолянкина

М о с к в а   1998

Предложения представили:

И.Г. Алексеев, В.С. Борисов, Л.С. Воробьев, А.Г. Долголенко,
В.П. Канавец, Ю.Т. Киселев, В.М. Колыбасов (ФИАН), М.В. Косов,
А.П. Крутенкова, В.В. Куликов, Г.А. Лексин, Н.А. Пивнюк, В.В. Рыльцов,
Д.Н. Свирида, В.А. Смирнитский, В.Т. Смолянкин, В.Л. Столин,
В.В. Сумачев (ПИЯФ), Ю.В. Требуховский, В.П. Чернышев,
В.А. Шейнкман.

# Оглавление

# 1. Введение

В настоящее время промежуточные энергии представляются особенно перспективным полем исследований для развития теории сильных взаимодействий. Это — область применения быстро развивающейся непертурбативной квантовой хромодинамики и связанных с ней моделей, опирающихся на базовые свойства этой теории. Дальнейший прогресс физики промежуточных энергий требует проведения новых прецизионных экспериментов в интервале (1–10) ГэВ. Об этом, например, убедительно свидетельствует ввод в действие новых исследовательских ускорителей COSY, CELSIUS в Европе, CEBAF в США, работающих в интервале (1–6) ГэВ. Синхротрон ИТЭФ У-10 обеспечивает экспериментальные установки протонными пучками в значительно более широком энергетическом интервале (1–10) ГэВ и, что особенно важно, пучками пионов с импульсами (0.5–5) ГэВ/$c$, сегодня доступными лишь на трех ускорителях мира: ИТЭФ, КЕК (Япония), BNL (США). Планируется также вывод ионных пучков с энергиями до 5 ГэВ/нуклон. Эти возможности У-10 позволяют исследовать ряд важных вопросов физики адрон-адронных, адрон-ядерных и ядро-ядерных взаимодействий.

На синхротроне ИТЭФ может быть выполнена большая программа исследований по адронной физике, требующая восстановления режима ускорения протонов до 8–10 ГэВ. Условно эти работы можно разделить на две группы: работы, максимально приближенные к условиям полного опыта, имеющие конечной целью парциально-волновой анализ, и работы поискового типа, направленные на открытие новых неизвестных распадов известных адронов и на поиски новых экзотических состояний, которые не укладываются в классификацию наивной кварковой модели. Оба эти направления дополняют друг друга и имеют общую цель — расширение существующей экспериментальной базы данных для развития теории сильных взаимодействий.

Сейчас достаточно надежно установлено, что основные мезоны и барионы составлены из кварков (и антикварков), а взаимодействие между ними переносится глюонным полем. Тем не менее, проблема конфайнмента еще не имеет решения на основе общих принципов теории. Для этой проблемы может оказаться решающим более детальное знание свойств глюонного и кваркового конденсатов. За последние десятилетия накоплен значительный экспериментальный и теоретический материал, не укладывающийся в наиболее простые представления о глюонных и кварковых полях. С использованием правил сумм ИТЭФ было установлено, что вакуум глюонных и кварковых полей не совпадает с вакуумом теории возмущений — существуют глюонный и кварковый конденсаты. Из этого можно заключить, что, наряду с адронными состояниями обычных типов $q\bar{q}$ и $qqq$, могут существовать нестандартные адроны, связанные с возбуждениями кваркового и глюонного конденсатов. Пока нет общепринятого развернутого теоретического описания непертурбативного вакуума, предсказать свойства таких нестандартных состояний затруднительно; они могут обладать необычными свойствами ро-ждения и распада и аномальной четностью. Кандидатами в такие нестандартные адроны являются хорошо установленные резонансы $f_0(980)$, $a_0(980)$ и, возможно, наблюденные недавно на шестиметровом спектрометре ИТЭФ узкие резонансы в $K_S^0 K_S^0$-системе. Один из них, пока надежно не подтвержденный, с предполагаемыми квантовыми числами $f_2(1245)$ может быть исследован на протонном синхротроне ИТЭФ. Область масс около 980 МэВ вполне доступна для исследований на синхротроне ИТЭФ и, по-видимому, еще не исчерпана. Кроме таких нестандартных мезонов, существует ряд кандидатов в глюболы, классификация которых пока не установлена; часть из них находится в области масс, доступной для исследования на синхротроне ИТЭФ. Экспериментальные и теоретические исследования глюониев, их смешивания с обычными $q\bar{q}$-состояниями, рождения и распадов были бы заметным вкладом в физику адронов.

Мезонная спектроскопия в интервале энергий ускорителя ИТЭФ по сравнению с более высокими энергиями обладает преимуществами больших сечений генерации, так как многие из неупругих процессов описываются однопионным или барионным обменами, амплитуды которых быстро уменьшаются с энергией. Во многих случаях измерение спиновых характеристик эксклюзивных процессов позволяет путем безмодельного анализа надежно разделить интенсивности участвующих волн.

Отдельной отраслью адронной физики является барионная спектроскопия. Интервалы импульсов пионных и каонных пучков ИТЭФ перекрывают всю область барионных резонансов $N$, $\Delta$, $\Lambda$, $\Sigma$. Для барионных состояний возможны реакции прямого рождения на $\pi$- и $K^-$-пучках. Особенно интересны исследования неупругих каналов, так как некоторые из них, например $\pi N \to \Lambda K$, $\pi N \to N\eta$, выделяют резонансы с изоспином 1/2.

Систематика, основанная на парциально-волновых анализах упругих и неупругих каналов, еще не исчерпана, и имеется обширное поле деятельности на синхротроне ИТЭФ при минимальной конкуренции. Особый интерес представляет изучение неупругих реакций с рождением $\phi$-мезонов (см. [1]), нацеленное на открытие резонансов со скрытой странностью. Интересны и реакции с рождением многих мезонов, для которых получены на 2-метровой камере ИТЭФ указания на каскадное рождение экзотического бариона с изоспином 5/2.

Весьма перспективной представляется и программа по адрон-ядерной и ядро-ядерной физике. Основные направления здесь связаны с изучением механизмов взаимодействия частиц в ядерной среде. Большое внимание привлекают исследования коллективных явлений в ядерных реакциях, таких как подпороговое рождение, рождение кумулятивных частиц, ненуклонные степени свободы в ядрах, адронизация кварков в ядерной среде. Интересны измерения пространственно-временных характеристик образования частиц в протон(ядро)-ядерных взаимодействиях. Все эти работы активно развивались в ИТЭФ, и были получены пионерские результаты. На стыке ядерной физики высоких энергий и резонансной физики стоят поиски мно-гонуклонных образований в свободном состоянии. Примером может быть $d'$-дибарион, существование которого исследовалось теоретиками и экспериментаторами в ИТЭФ. Предлагается также проверка предсказываемой модификации свойств адронных резонансов в ядерной среде. Могут быть получены данные о вызывающих в последнее время большой интерес состояниях с большой фазовой плотностью, которые в иных условиях предлагается исследовать на установке ALICE (LHC).

Представленная ниже предварительная программа, разработанная экспериментальными лабораториями и группами, охватывает лишь часть перечисленных возможностей. При соответствующей моральной и материальной поддержке может быть разработана более широкая программа экспериментов на синхротроне ИТЭФ.

## 2.    Адрон-адронные процессы

### 2.1.    Поиски и исследование экзотических состояний

Как следует из кварковой модели, все известные мезоны являются связанными состояниями пары кварк-антикварк $(q\bar{q})$, а барионы — трех кварков. Однако ни одна теоретическая модель не закрывает возможность существования более сложных объектов типа многокварковых $qq\bar{q}\bar{q}$-мезонов, $qqqq\bar{q}$-барионов и $qqqqqq$-дибарионов, а также гибридов и глюболов. Поиски таких нестандартных экзотических образований являются важнейшей задачей адронной спектроскопии в настоящее время.

Адронные экзотические состояния могут быть трех видов.

1. С явной (открытой) экзотикой — имеющие экзотические главные квантовые числа (мезоны с зарядом или странностью $\pm2$, изоспином $> 1$, барионы с зарядом $> 2$ и т.д.). Такие объекты должны быть многокварковыми.

2. Образования с экзотическим набором квантовых чисел $J$, $P$, $C$, например, мезоны с $J^{PC} = 0^{+-}, 0^{--}, 1^{-+}$ и т.д.

3. Состояния со скрытой экзотикой, отличающиеся аномальными динамическими свойствами: аномально малой шириной, необычными каналами распада, специальными каналами образования и т.д.

Следует заметить, что до сих пор нет надежно установленных экзотических адронов, хотя можно назвать несколько серьезных кандидатов, которые могут считаться таковыми [2]. Наличие на ускорителе ИТЭФ интенсивных пучков $\pi^-$-мезонов с импульсами в диапазоне $3 \div 5$ ГэВ/$c$ и протонов с импульсами в диапазоне $3 \div 8$ ГэВ/$c$ предоставляет хорошую возможность провести детальное изучение экзотических состояний всех трех видов.

Как показывают последние эксперименты (см., например, [3]), ожидаемые сечения образования экзотических состояний лежат в нанобарновой области.

Массы экзотических барионных состояний, таких как $E_{55}$, $\Delta^{++}\pi^+$, $p\varphi$, $p\eta$, $\Sigma K$, $\Sigma^* K$ и др., лежат в диапазоне $1.5 \div 3$ ГэВ, доступном на ускорителе ИТЭФ. Продукты распада таких тяжелых резонансов будут вылетать в широком телесном угле. Поиски скрытой экзотики предполагают изучение, в первую очередь, узких резонансов. Все эти обстоятельства требуют использования интенсивных первичных пучков ($\sim 10^6$ частиц/сек.) и широкоапертурного (близкого к $4\pi$) прецизионного ($\Delta m = 5 \div 10$ МэВ) спектрометра с возможностью регистрации заряженных и нейтральных частиц.

Рассмотрим некоторые процессы, в которых можно ожидать образования экзотических состояний, и соответствующие методы исследования.

**2.1.1.** *Сканирование по энергии начальной частицы сечений эксклюзивных каналов реакций,* например:

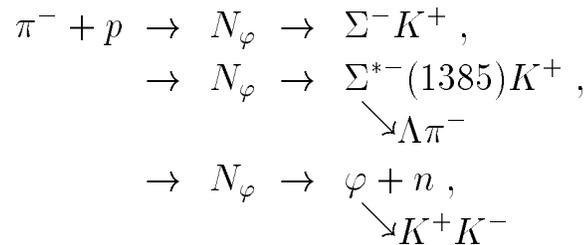

позволит провести поиски экзотического бариона в прямом канале.

**2.1.2.** *Прецизионное измерение спектров эффективных масс в реакциях*

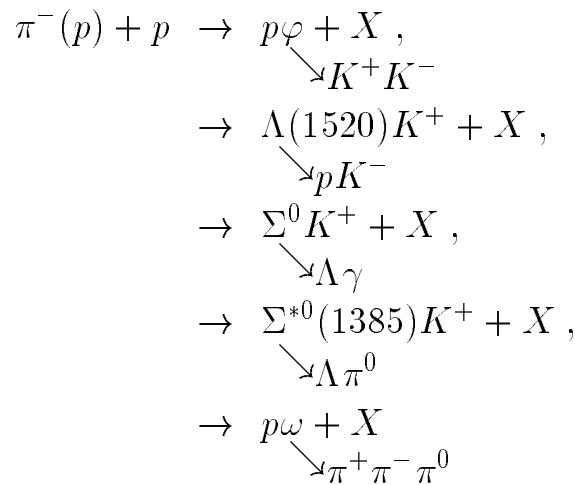

и др. дает возможность провести поиски узких барионов со скрытой экзотикой. Указания на существование экзотических состояний в системах $\Sigma^0 K^+$ и $\Sigma^{*0}(1385)K^+$ были получены в работе [4].

**2.1.3.** *Поиски экзотических мезонов в процессах с барионным обменом.*

Можно ожидать, что возбуждение внутренних степеней свободы, приводящих к образованию экзотических состояний, будет происходить при больших переданных импульсах, например, в реакциях с барионным обменом. Наиболее удобно, с точки зрения эксперимента, изучать такие процессы в $pp$-, а не в $\pi^- p$-реакциях: в этом случае экзотические мезоны будут вылетать вперед и захватываться спектрометром с широкой апертурой. Особый интерес представляет поиск экзотического мезона с зарядом +2 в реакции

$$p + p \to M^{++} + [p\pi^- n] \qquad (\Delta^- \text{-обмен}).$$

Известно, что сечения с барионным обменом быстро падают с энергией ($E^{-n}$, где $n \approx 3 \div 5$ для энергий $2 \div 50$ ГэВ), поэтому диапазон энергий, доступный на ускорителе ИТЭФ, чрезвычайно выгоден для исследования такого рода процессов.

**2.1.4.** *Поиски барионного резонанса $E_{55}$.*

Как мы упоминали выше, ряд теоретических моделей предсказывает существование экзотических барионов. В частности, применение метода правил сверхсходящихся сумм к рассеянию реджеонов со спином $J = 1$ ($\pi$, $\rho$) на нуклонах и $\Delta_{33}$-изобаре [5] привело к предсказанию существования серии экзотических барионов с изоспином $I \geq 5/2$, спином $J = 1$ и положительной P-четностью. Самый легкий член этого семейства экзотических барионов имеет квантовые числа $J = 5/2$, $I = 5/2$ и достаточно маленькую массу $M = 1.4 \div 1.7$ ГэВ.

$E_{55}$-барион является аналогом $N$ и $\Delta_{33}$ в мире резонансов с $I = 5/2$; это — основное состояние среди этих резонансов. Наиболее вероятный канал распада $E_{55}$ есть каскадный процесс $E_{55} \to \Delta_{33}\pi \to N\pi\pi$. При малых импульсах распада $E_{55} \to \Delta_{33}\pi$ для различных масс $M_E$ предсказаны ширины $\Gamma_E$ [6], значения которых приведены в таблице.

| $M_E$, ГэВ | 1.42 | 1.44 | 1.52 | 1.56 | 1.60 | 1.65 |
|---|---|---|---|---|---|---|
| $\Gamma_{E \to \Delta\pi}$, МэВ | 21 | 36 | 140 | 215 | 307 | 443 |

Резюмируя данный раздел, можно сказать, что нужно детально исследовать резонансы в $p\pi^+\pi^+$-системе в области масс $M_{\text{порог}} < M_{p\pi^+\pi^+} < 1.8$ ГэВ, наложив ограничения на массу $p\pi^+$-подсистемы в области резонанса $\Delta(1232)$.

Первые эксперименты по исследованию $p\pi^+\pi^+$-системы были проведены в начале шестидесятых годов; в частности, этому вопросу была посвящена работа [7], в которой представлены результаты изучения реакции $\pi^+ p \to \pi^- \pi^+ \pi^+ p$ при $p_{\pi^+} = 3.65$ ГэВ/$c$.

В дальнейшем появился ряд работ, посвященных поискам $E_{55}$-бариона (см. работу [8] и ссылки в ней). В ИТЭФ этой проблемой занимался ряд экспериментальных групп, использовавших как методику пузырьковых камер, так и электронную аппаратуру [5, 9, 10, 11].

Следует подчеркнуть, что экспериментальная ситуация с $E_{55}$ крайне противоречива. В одних работах сообщается о наблюдении $E_{55}$ и приводятся его массы, ширины и сечения образования. В других работах $E_{55}$ не наблюдается, и дается верхний предел на сечение его образования. Кроме того, в работах, где сообщается о наблюдении этого бариона, существуют противоречия по положению масс и ширин.

Наиболее эффективно искать $E^{+++}$-резонансы в процессах на $\pi$-мезонном пучке в эксклюзивной реакции

$$\pi^+ p \to p\pi^+\pi^+\pi^- \tag{1}$$

в кинематике, в которой $p\pi^+\pi^+$-система летит по направлению пучка. Выбор такой кинематики может быть осуществлен с помощью простого триггера, когда $\pi^-$ в лабораторной системе вылетает назад. Диаграмма, описывающая этот процесс, соответствует обмену в $u$-канале $\Delta$-изобарой (см. рис. 1).

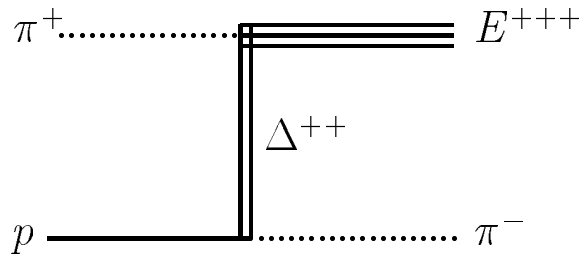

Рис. 1: Обмен $\Delta$-изобарой.

В этой кинематике важно то, что фоновые процессы, отражающиеся на $\Delta\pi$-системе, также происходят за счет барионного обмена. Благодаря этому, отношение вероятностей выходов фоновых систем и резонансов может быть не таким большим, как в области протонной фрагментации. На рис. 2 показана диаграмма фонового процесса $\pi^+ p \to \Delta^{++} V^0$.

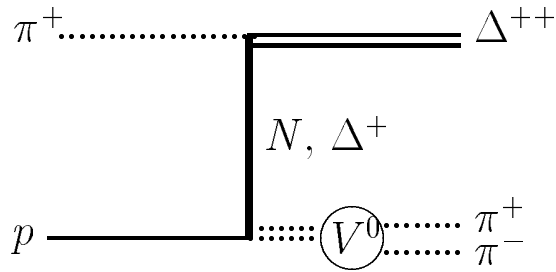

Рис. 2: Диаграмма процесса $\pi^+ p \to \Delta^{++} V^0$.

Кроме реакции (1) представляет интерес реакция $\pi^+ p \to p\pi^+\pi^+\pi^-\pi^0$, в которой можно искать совместное рождение $E^{+++}$ и $\rho$, идущая согласно диаграмме рис. 3. Существует интересная возможность поиска $E_{55}$-резонансов в реакциях

$$
\begin{aligned}
pp &\to \pi_f^- p\pi^+\pi^+ n\,, \\
pp &\to n_f p\pi^+\pi^+\pi^-
\end{aligned}
\tag{2}
$$

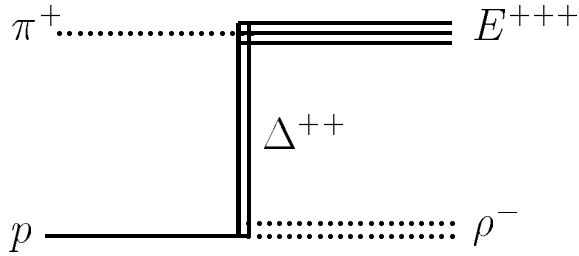

Рис. 3: Диаграмма совместного рождения $E^{+++}$ и $\rho$.

при специальной конфигурации импульсов, имитирующих процесс рассеяния назад. Диаграммы для этих процессов показаны на рис. 4.

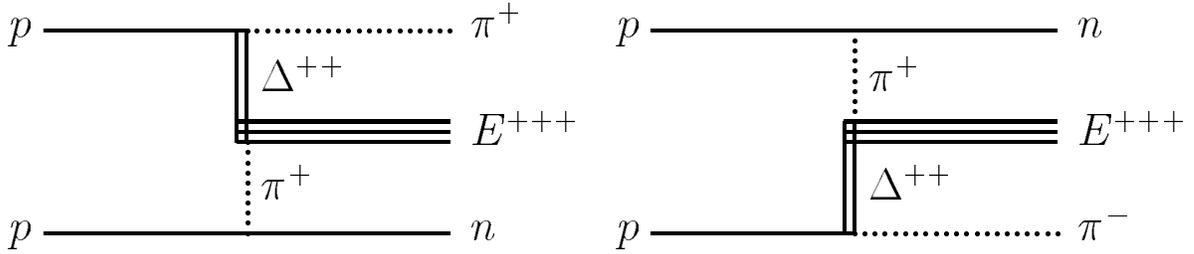

Рис. 4: Диаграммы процессов рождения $E_{55}$ в $pp$-взаимодействии.

Так как множественность заряженных частиц в реакции (2) меньше, чем в реакции $pp \to \pi_f^- p\pi^+\pi^- p$, то первая реакция становится более предпочтительной из-за того, что число фоновых процессов для нее меньше.

Одним из наиболее прямых способов прояснения экспериментальной ситуации с возможным рождением состояния $E^{+++}$ может быть набор значительного статистического материала по реакциям $\pi^+ p \to \Delta^{++} + \pi^+\pi^-$ и $\pi^+ p \to \Delta^{++} + \pi^+\pi^-\pi^0$ при импульсе первичного $\pi^+$ от 3 до 4 ГэВ/$c$ и, возможно, реакций с большим числом вторичных пионов. Именно в этих реакциях на ограниченном статистическом материале двухметровой жидководородной камеры ИТЭФ были получены указания на рождение состояния $E^{+++}$ в процессах каскадного распада тяжелых $\Delta$-изобар: $\pi^+ p \to \Delta^{*++} \to E^{+++} + \pi^-(\rho^-)$. Такой процесс $s$-канального типа не имеет малости, присущей реакциям барионного обмена, и мог бы быть исследован на многочастичном спектрометре.

**2.1.5.** *Поиски экзотического дибарионного резонанса $d'$ ($M \approx 2.06$ ГэВ, $\Gamma < 10$ МэВ) в реакции $pd \to pd' \to ppp\pi^-$.*

Дибарионный резонанс $d'$ с открытой экзотикой может рассматриваться как многокварковый мешок в свободном состоянии. Сравнение характеристик многокварковых мешков в ядрах (флуктоны) и в свободном состоянии позволяют в принципе судить о радиусе конфайнмента в ва-кууме и ядерной среде. Радиус конфайнмента в ядерной материи, несомненно, важнейшая характеристика квантовой хромодинамики сред.

В ряде теоретических работ [12] предсказывается существование $d'$ как многокваркового мешка. Наблюдение $d'$ будет тестом справедливости этих теоретических подходов.

Существует ряд статистически достоверных явлений (например, резонансное поведение сечения "упругой" двойной перезарядки пионов [13]), которые можно трактовать, как существование ядерных систем, включающих дибарион $d'$.

Имеются экспериментальные указания на существование $d'$ в свободном состоянии, первое из которых было получено в ИТЭФ [14]. Необходимо набрать бо́льшую статистику и подтвердить или опровергнуть существование свободного $d'$. Если $d'$ есть, то существуют $d'$- ядра, если нет, то надо искать альтернативные объяснения безусловно наблюденных аномалий в образовании пионов с энергией 50 МэВ.

Обнаружение $d'$ и выделение соответствующих событий позволят в перспективе оценить размер $d'$ стандартным методом $pp$-корреляций и, следовательно, его плотность и сравнить с плотностью флуктонов.

*Свойства $d'$.* Из прямого эксперимента и анализа данных по перезарядке на ядрах известна следующая информация о $d'$:

а) $M_{d'} \approx 2.06$ ГэВ;

б) $d' \to pp\pi^-$ или $nn\pi^+$, $I = 0$, $J^P = 0^-$;

в) энерговыделение $\sim 50$ МэВ;

г) по-видимому, образуется при взаимодействии с тесными группами нуклонов в ядрах или на малых расстояниях в $pp$-взаимодействиях;

д) в $pp$-взаимодействиях $d'$ удалось выделить лишь среди событий, для которых масса системы $pp$ $M_{pp} - 2m_p < 18$ МэВ – узкие $pp$-пары в конце реакции;

е) $\Gamma_{d'}^{\text{эксп}} < 10$ МэВ и, по-видимому, $\leq 1$ МэВ; экспериментальная ширина обусловлена разрешением установки;

ж) узость резонанса, возможно, связана с его строением — он представляется [15] квадрокварк-дикварковым мешком, что хорошо согласуется с характеристикой, указанной в п. д).

*Выбор реакции, кинематика.*

I. Предлагается искать $d'$ в квазибарионной реакции $pd \to pd'$; $d'$ — очень узкий резонанс (п. е)).

II. Отбирать такие события реакции

$$pd \to pd' \to ppp\pi^- ,\tag{3}$$

в которых $p$ летит назад в л.с. Это, конечно, жесткий отбор, но он выделяет события, в которых налетающий протон взаимодействует с $d$ в "сжатом" состоянии, как при обратном $pd$-рассеянии

$$pd \to dp \ . \tag{4}$$

Таковы условия рождения $d'$ (см. п.п. г), д), ж)).

III. Реакция $pd \to dp$ быстро вымирает с начальной энергией, но ее сечение вполне измеримо в диапазоне энергий от порога пионного рождения до нескольких ГэВ. Угловое распределение протонов позволяет предполагать вклад диаграммы рис. 5 с обменом нуклоном, что обеспечивает малость расстояний при взаимодействии.

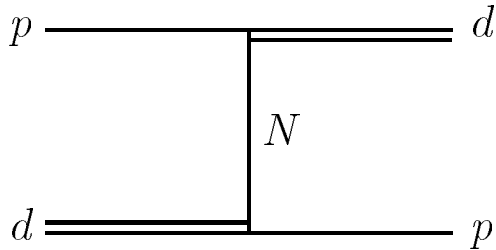

Рис. 5: Диаграмма нуклонного обмена.

IV. Реакция (3) может идти согласно диаграмме рис. 6 с обменом $N^*$,

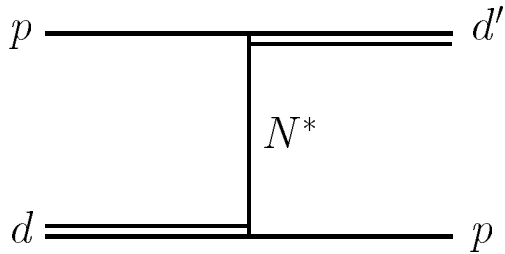

Рис. 6: Диаграмма обмена $N^*$.

что только уменьшает межнуклонное расстояние при взаимодействии и способствует перестройке $6q$-мешка. Вероятность реакции (3), конечно, меньше чем (4), но из-за обмена изобарой — не слишком: сравнить обратное рассеяние $\pi^+ p$ (нуклонный полюс) и $\pi^- p$ (в пропагаторе изобара).

V. В случае детектирования вылета протона назад кинематика $d'$ жестко задана. Возможен анализ по методу недостающих масс, но предполагается регистрировать **все** продукты распада $d' \to pp\pi^-$ и видеть его по инвариантной массе на выделенной части фазового объема.

VI. $d'$ — маловозбужденное состояние, система двух протонов летит по направлению $d'$ (при начальной энергии 1 ГэВ отклонение $< 5^o$).

Так как необходимо отбирать "узкие" пары протонов (см. п. д)), то и протоны летят по направлению $d'$. Это те протоны, которые в системе $d'$ летят назад.

VII. Пионы от распада $d'$, летящие в его системе вперёд, отходят от направления $d'$ на углы до $30^o$ и имеют при $T \sim 1$ ГэВ до $300 \div 400$ МэВ/$c$.

IX. Все частицы могут быть зарегистрированы счётчиками системы КОРД ("КОРреляционный Детектор" [16]). Предыдущий эксперимент проводился с системой БАС ("Безмагнитный Адронный Спектрометр"). КОРД является усовершенствованным вариантом последней.

*Постановка опыта* вытекает из расположения счётчиков КОРД.

Начальная энергия определяется следующими соображениями:

* вылетающий назад протон должен иметь достаточный импульс для регистрации системой КОРД (1-й слой), при $T_{\text{нач.}} = 1$ ГэВ $\quad p_p \approx 250 \div 350$ МэВ/$c$;

* следует отойти вверх по энергии от порога образования $d'$ для увеличения сечения;

* превышение порога должно быть небольшим, чтобы подавить фоны, связанные с рождением других пионов.

Предполагается работа на жидкодейтериевой мишени, чтобы избежать разностного опыта типа $C^2H_2 - C$. Чтобы избежать опыта $(d + Мишень) - Мишень$ надо иметь систему координатных детекторов, выделяющих в простой геометрии случаи на дейтерии в мишени. Это эффективно поможет увеличению статистической точности.

**2.1.6.** *Образование экзотических состояний в реакциях однопионного обмена.*

Недавно обнаруженные мезоны, являющиеся кандидатами в экзотические состояния, например, $g_T(2010)$, $G(1590)$ и др., образуются в зарядово-обменной реакции $\pi^- p \to M^0 n$, в которой доминирует однопионный обмен. Эти эксперименты проведены при импульсе $\pi^-$-мезонов $25 \div 32$ ГэВ/$c$. Поскольку сечение таких реакций падает с ростом энергии как $E_\pi^{-2}$, исследование этих и поиск других состояний могут быть проведены на ускорителе ИТЭФ с существенно большей статистической обеспеченностью, чем в предыдущих работах. Например, число событий, которые можно получить на широкоапертурном спектрометре за 100 часов работы ускорителя ИТЭФ для реакции $\pi^- p \to \omega n$ (сечение реакции $\approx 200$ мкб), составит $\approx 10^8$, а для реакции $\pi^- p \to \eta' n$ — $\approx 10^7$ (сечение реакции $\approx 30$ мкб).

В ряде случаев при работе на пионном пучке можно будет провести парциально-волновой анализ, который существенно увеличит возможности и надежность изучения экзотических адронных состояний.

## 2.2. Исследование низколежащих скалярных мезонов

Предлагается исследование реакций

$$\pi^- p_\uparrow \;\rightarrow\; \pi^+ \pi^- n \;, \tag{5}$$

$$\pi^- p_\uparrow \;\rightarrow\; \pi^0 \pi^0 n \;, \tag{6}$$

$$\pi^- p_\uparrow \;\rightarrow\; K^+ K^- n \tag{7}$$

на поляризованной протонной и жидководородной мишенях в интервале начальных импульсов (2.5–4.0) ГэВ/$c$, обеспечивающем перекрытие областей скалярных резонансов $f_0(750)$, $f_0(980)$, $a_0(980)$ и $f_0(1370)$. Эта программа является продолжением и расширением эксперимента по изучению генерации пионов пионами на поляризованных протонах, выполненного в последние годы на синхротроне ИТЭФ [17].

Основная цель предполагаемых экспериментов состоит в исследовании скалярного сектора бозонной спектроскопии. Известно, что до настоящего времени не установлен нонет скалярных мезонов $J^{PC} = 0^{++}$, хотя и заполнены нонеты псевдоскалярных, векторных и тензорных мезонов. Для скалярного нонета имеются кандидаты с изоспином 1: $a_0(980)$ и $a_0(1450)$, и с изоспином 0: $f_0(400 \div 1200)$, $f_0(980)$ и $f_0(1370)$, но некоторые из них проблематичны, так как сравнительно узкие ($\Gamma = 40 \div 100$ МэВ) $f_0(980)$ и $a_0(980)$ могут оказаться связанными $K\bar{K}$-состояниями или "новыми адронами" [18]. С другой стороны, поиск экзотических состояний (4-кварковых ($q\bar{q}q\bar{q}$) мезонов, гибридных ($q\bar{q}g$) мезонов и глюболов ($gg$)) особенно перспективен в скалярном секторе, так как можно ожидать, что низшие состояния этих объектов являются скалярами. Особый интерес представляет сигма-мезон $f_0(400 \div 1200)$, имеющий длинную и противоречивую историю. На существование этого мезона указывает ряд теоретических работ и экспериментальных результатов. Так, в модели "башен", состоящих из вырожденных резонансов с чередующейся четностью [19], $\sigma$-мезон с $J^{PC} = 0^{++}$ возникает как член первой башни ($m_\sigma = m_\rho$), принадлежащий дочерней траектории Редже, смещенной в $j$-плоскости вниз на единицу относительно обменно-вырожденной лидирующей траектории с $\sigma \cdot P = 1$. Существование таких "башен" следовало из отсутствия $u$-канальных резонансов (и пика назад) в реакциях $\pi^+ \pi^- \rightarrow \pi^+ \pi^-$. Имеющиеся экспериментальные данные по мезонным резонансам не противоречат существованию прямолинейных дочерних траекторий, естественно возникающих в струнных моделях, по крайней мере в области масс выше 1 ГэВ [20]. Полевая модель Намбу-Иона-Ласинио [21] предсказывает $m_\sigma = 2m_q$, где $m_q$ — масса конституентного кварка. Современные полевые модели, удовлетворяющие киральной инвариантности КХД (см., например, работу Дельборго и Скадрона [22]), предсказывают сигма-мезон с $m_\sigma = 2m_q \approx 650$ МэВ.

Экспериментальные данные по реакции (5), полученные при 1.78, 6, 12 и 17 ГэВ/$c$, свидетельствуют о сложной структуре интенсивности $S$-волны, содержащей, в частности, пик при массе $\sim 750$ МэВ с шириной (100–200) МэВ. Поведение относительной фазы $S$- и $P$-волн также соответствует существованию сравнительно узкого резонанса. В то же время стандартная АМР-параметризация упругого $\pi\pi$-рассеяния противоречит наличию узкого резонанса, хотя и может быть согласована с широким (Г = 400 ÷ 600 МэВ) сигма-мезоном.

Дальнейший прогресс в изучении скалярных мезонов требует хорошо статистически обеспеченных измерений матричных элементов реакций (5), (6), (7). Во всех этих реакциях выделяются бозонные резонансы с натуральной спин-четностью, в реакцию (6) вносят вклад лишь резонансы с четным спином, реакцией (7) выделяются бозонные резонансы с большой вероятностью распада по каналу $K\bar{K}$. В обсуждаемой энергетической области нет данных о спиновой зависимости реакции (6). Отсутствуют также сведения о спиновой зависимости реакции (7) в области $f_0(980)$.

Исследование процессов (5), (6), (7) обеспечит прямое модельно-независимое восстановление интенсивностей и относительных фаз участвующих амплитуд. Ожидаемые результаты так же принесут ценную информацию о динамике этих реакций, включая энергетическую зависимость матричных элементов и амплитуд и оценки вкладов "не-пионных" механизмов обмена. Они также позволят проверить модели спиновой зависимости механизмов генерации дипионов (дикаонов) пионами (например, Кимель-Оуэнс [23]).

При реализации экспериментов может быть использован опыт разработки и эксплуатации поляризованных мишеней в ИТЭФ, имеющиеся криовакуумная инфраструктура установки СПИН и пакеты программ восстановления матричных элементов и амплитудного анализа.

## 2.3. Измерение параметров вращения спина при промежуточных энергиях и барионная спектроскопия

Барионная спектроскопия является одной из основ современной физики сильных взаимодействий при промежуточных энергиях. Разнообразные модели взаимодействия конституэнтов, включая струнные, двухчастичные релятивистские и нерелятивистские потенциальные модели, модели мешков, скирмионы и правила сумм КХД [24], нуждаются в детальной информации о массах, ширинах и модах распада резонансов. Некоторые из этих моделей довольно успешно описывают спектр и ряд характеристик адронов через конституентные кварки. Но связи этих моделей с основополагающей теорией КХД недостаточно обоснованы.

Существуют принципиально важные открытые вопросы барионной спек-троскопии, в частности, о роли глюонных степеней свободы при низких энергиях. Например, несмотря на тот факт, что КХД не запрещает существование гибридных ($qqqg$) резонансов, ни один из таких резонансов не установлен экспериментально. Имеют место и загадочные явления существования кластеров резонансов, имеющих почти одинаковые массы, но разные спины. Частичное объяснение этого эффекта, возможно, содержится в существовании дублетов резонансов, вырожденных по четности. Последнее явление может быть объяснено как следствие прямолинейности траекторий Редже [25] или как проявление приблизительной киральной симметрии КХД [26].

В области начальных импульсов до 2 ГэВ/$c$ состояние экспериментальной барионной спектроскопии требует существенного улучшения. В настоящее время в указанном диапазоне известно три группы парциально-волновых анализов: **CMB80** [27], **KH80** и **KA84** [28], **SM90**, **SM95**, **SP98** и другие анализы **VPI** [29]. В табл. 1 приведены характеристики $\Delta$-резонансов во второй резонансной области по результатам этих трех фазовых анализов. Действительно, по данным PDG [30] из 20 наблюдаемых резонансов только половину можно считать твердо установленными. Более того, предсказания различных парциально-волновых анализов (ПВА) заметно различаются как в отношении спектра и свойств резонансов, так и в экспериментальных наблюдаемых. Возникают даже вопросы относительно существования казалось бы ранее твердо установленных резонансов. Так, в недавней работе Г. Хелер привел доказательство отсутствия $S_{11}(1535)$ [31].

Таблица 1: Резонансы $I = 3/2$ во второй резонансной области.

|  | Статус по PDG | KH80 | CMB | VPI(SM90) |
|---|---|---|---|---|
| $S_{31}(1990)$ | *** | 1908(140) | 1890(170) | нет |
| $P_{31}(1910)$ | **** | 1888(280) | 1910(225) | 1950(400) |
| $P_{33}(1920)$ | *** | 1868(200) | 1920(200) | нет |
| $D_{33}(1940)$ | * | нет | 1940(200) | нет |
| $D_{35}(1930)$ | *** | 1901(195) | 1940(320) | 2018(400) |
| $F_{35}(1905)$ | **** | 1905(260) | 1910(400) | 1794(230) |
| $F_{37}(1950)$ | **** | 1913(224) | 1950(340) | 1884(240) |

В последние годы в связи с пуском нового ускорителя CEBAF (Jefferson Laboratory) и принятой на электронных ускорителях программах развития барионной спектроскопии с использованием электромагнитных пробников повысился и интерес к аналогичным работам с адронными пробниками. Сведения о параметрах резонансов на пионных (каонных) пучках необходимы для интерпретации результатов работ по электро- и фоторождению барионных резонансов и полноты описания их распадов.

Эта работа представляет собой прямое продолжение работы на установке СПИН-ЛМ по измерению параметров вращения спина [32, 33, 34].

Измерения параметров вращения спина, представляющие собой трудный "двухспиновый" эксперимент, до проведенных в ИТЭФ опытов были реализованы только лабораториями ПИЯФ [35] и LAMPF [36] в первой резонансной области при импульсах, меньших 750 МэВ/$c$, т.е. лишь в первой трети ре-зонансной области. Информация о параметрах вращения спина дополняет данные по дифференциальным сечениям и поляризации до "полного опыта", позволяющего осуществить прямое восстановление участвующих амплитуд. Отсутствие этих данных приводит к так называемым дискретным (Бареллетовским) неоднозначностям фазового анализа и является потенциально опасным для правильного выбора истинных решений ПВА. Принятые в настоящее время спектр и характеристики резонансов опираются, в основном, на результаты весьма старых анализов **СМВ** и **КН**, не включающих данных по параметрам вращения спина. Одновременно с этим анализы группы **VPI**, используя более новые и точные данные, предсказывают заметно отличные параметры резонансов и находятся в явном противоречии с **СМВ** и **КН** в предсказаниях по параметрам вращения спина, особенно ярко выраженных во второй резонансной области.

Недавние измерения параметров вращения $A$ и $R$ [32] были проведены в ИТЭФ в сотрудничестве с ПИЯФ при начальном импульсе 1.43 ГэВ/$c$ для упругого $\pi^+p$-рассеяния. Этот импульс соответствует положению кластера семи почти вырожденных изобар с массой, близкой к 1.9 ГэВ. Результаты эксперимента оказались неожиданными: данные бесспорно подтвердили предсказания **VPI** (рис. 7). Дальнейший анализ с применением метода

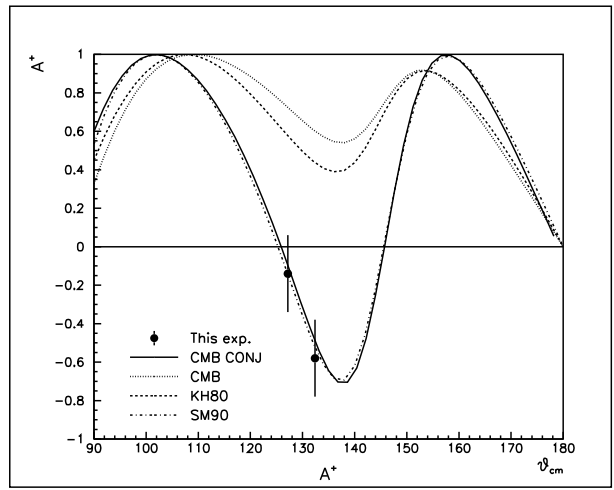

Рис. 7: Параметр вращения спина $A$ в реакции упругого рассеяния $\pi^+p \to \pi^+p$ при $P_{\pi^+} = 1.43$ ГэВ/$c$ как функция угла рассеяния в СЦМ. Кривые — предсказания фазовых анализов. Экспериментальные результаты получены на ускорителе ИТЭФ.

нулей поперечных амплитуд (метод Барелле) [37] показал, что указанные разногласия являются проявлением Бареллетовской неоднозначности, приведшей к выбору неправильной ветви решения анализами **СМВ** и **КН** в точке ветвления около 1 ГэВ/$c$. Развитие метода траекторий нулей поперечных амплитуд позволило внести коррекцию в существующие фазовые анализы с учетом полученных данных, приводящую к устранению их разногласий по параметрам вращения спина в широком энергетическом диапазоне (СМВ CONJ на рис. 7 соответствует скорректированному анализу **СМВ**).

Анализ диаграмм Аргана показал, что согласие предсказаний фазовых анализов **VPI** со скорректированными анализами групп **KH** и **CMB** существенно улучшается для хорошо выраженных резонансов (см. рис. 8б). В то же время возникают существенные изменения параметров резонансов с рейтингом три звезды, сильно уменьшается упругость (см. рис. 8а); согласие в поведении диаграмм Аргана улучшается и в этом случае; это, возможно, объясняет отсутствие некоторых из резонансов в результатах анализа **VPI** (см. табл. 1 выше).

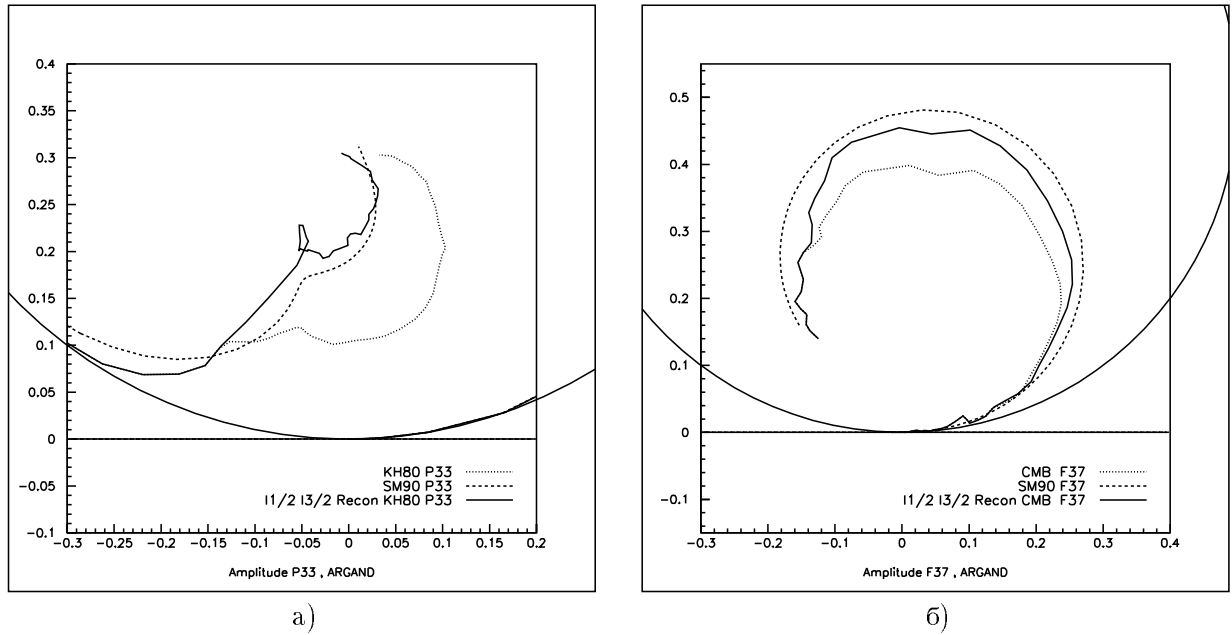

Рис. 8: Диаграммы Аргана для волн $P_{33}$ (а) и $F_{37}$ (б). Сплошная кривая на обоих рисунках обозначает скорректированный анализ.

Предлагается продолжить измерения параметров вращения в ряде других кинематических областей в диапазоне $1 \div 2$ ГэВ/$c$, в которых либо имеются серьезные разногласия в предсказаниях существующих фазовых анализов, либо существуют подозрения на двузначность решения ПВА, полученные путем анализа их поведения методом Барелле. Этот метод, получивший дальнейшее развитие на первом этапе работы, позволяет эффективно планировать выбор областей измерений.

Следует особо подчеркнуть, что существующие пионные пучки ускорителя У-10 обеспечивают возможность эффективного развития работ по измерениям спиновых наблюдаемых для барионной спектроскопии. Ускоритель ИТЭФ вполне конкурентоспособен по сравнению с другими ускорителями в мире по качеству и интенсивности пионных пучков.

Эксперимент проводится в тесном сотрудничестве с Петербургским институтом ядерной физики. Мы рассчитываем также на продолжение совместной работы с группой **VPI** (Вирджиния, США) по интерпретации результатов.

## 2.4. Исследование структуры $\eta'$- и $f_0$-мезонов и изучение механизма мезон-нуклонного взаимодействия

Целью эксперимента является измерение формы спектра недостающих масс в реакции рождения мезонов, а также относительных ширин их распадов для выяснения кварковой структуры легкого скалярного мезона $f_0(980)$ и проверки теоретических представлений о механизме взаимодействия $f_0(980)$- и $\eta'(958)$-мезонов с нуклонами. Сектор скалярных мезонов играет очень важную роль в физике адронов. Однако структура легких скалярных мезонов $a_0(980)$ и $f_0(980)$ не вполне ясна. Они могут быть $q\bar{q}$- или $qq\bar{q}\bar{q}$-состояниями, "молекулами" или другими нестандартными состояниями.

Такого рода измерения и эксперименты с $\eta'$- и $f_0$-мезонами пока еще не выполнены нигде. Часть планируемых измерений может быть выполнена на ускорителе COSY, но в существенно более узких рамках, т.к. максимальная энергия протонного пучка COSY всего лишь 2.6 ГэВ. Кроме того, обсуждаемые здесь физические задачи не вошли до сих пор в исследовательские программы на COSY.

Предлагается использовать внутреннюю водородную (дейтериевую) мишень из замороженных микрокапель газа (pellet target). Чтобы выделить тип и характер механизма взаимодействий $\eta'$- и $f_0$-мезонов с нуклонами, предлагается изучать их в определенных изоспиновых состояниях. Для этого предполагается измерять околопороговое поведение сечений реакций образования мезонов в следующих реакциях:

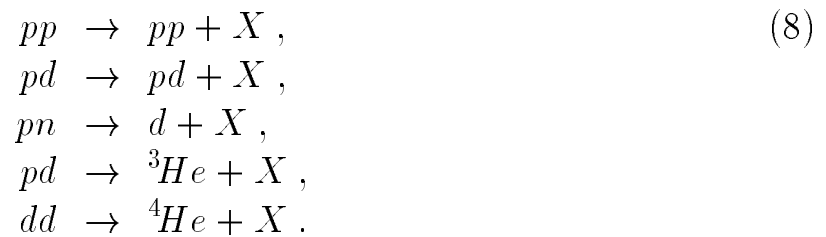

$$
\begin{aligned}
pp &\rightarrow pp + X\,, \\
pd &\rightarrow pd + X\,, \\
pn &\rightarrow d + X\,, \\
pd &\rightarrow {}^{3}He + X\,, \\
dd &\rightarrow {}^{4}He + X\,.
\end{aligned}
\tag{8}
$$

На второй стадии эксперимента предполагается измерять ширины электромагнитных распадов мезонов с помощью быстрого $LXe/LKr$-калориметра:

$$
\begin{aligned}
\eta' &\rightarrow \pi^0\pi^0\eta & (20.8 \pm 1.3)\%, \\
\eta' &\rightarrow 2\gamma & (2.12 \pm 0.13)\%.
\end{aligned}
$$

Улучшение точности измерения относительных ширин примерно на порядок прояснит две проблемы:

- содержания глюонной компоненты и доли странных кварков в $\eta'$;
- смешивания $\eta$- и $\eta'$-мезонов.

Измерения ширины распада

$$
f_0 \rightarrow \pi^0\pi^0 \quad (\sim 30\%)
$$

помогут прояснить кварковую структуру $f_0$-мезона.

Внутренняя мишень pellet target должна быть установлена в одном из прямолинейных промежутков нового трехгэвного ускорителя (накопительного кольца) ИТЭФ. Новый ускоритель имеет исключительно хорошие характеристики для проведения прецизионных опытов по исследованию адронных спектров в районе 1 ГэВ. Отметим лишь некоторые из них:

– высокая монохроматичность пучка протонов (без внешнего охлаждения) — $\Delta p/p \sim 5 \cdot 10^{-4}$;

– возможность получения пучков малого диаметра < 3 мм;

– сверхвысокий вакуум в кольце, снижающий общий фон;

– наличие больших прямолинейных промежутков, позволяющих свободно размещать детекторы;

– возможность изменять в широких пределах энергию пучка протонов, циркулирующего в кольце ускорителя.

Предлагается выполнять всю программу измерений в два этапа:

а) относительно дешевый вариант с использованием поворачивающего магнита самого ускорителя в качестве магнита спектрометра;

б) чтобы повысить точность измерений, следует увеличить аксептанс установки, используя магнит новой спектрометрической системы.

Для реализации первого этапа эксперимента мы предлагаем расположить pellet target перед поворотным магнитом ускорителя. Заряженные частицы, вылетающие из точки взаимодействия, отклоняются магнитным полем ускорителя. Поскольку точка взаимодействия пространственно хорошо определена, а поперечный импульс протонного пучка мал, то импульс вылетевших частиц может быть определен достаточно хорошо. Наличие сравнительно большой пролетной базы — до 3-4 метров позволит достаточно надежно идентифицировать частицы. Структура такой установки подобна установке COSY-11, но преимущества предлагаемой состоят в большем аксептансе спектрометра и бо́льшей энергии пучка.

Масса резонанса (например, в реакции (8)) может быть измерена, как недостающая масса к двум протонам, вылетевшим вперед. Импульс и $dE/dx$ этих протонов предполагается измерять передним детектором, который должен состоять минимум из шести плоскостей пропорциональных камер и $TOF - dE/dx$ счетчиковой детектирующей системы. Прототипы этой аппаратуры реализованы на установках COSY-11 и ANKE. Предполагается одновременно измерять другие вылетающие заряженные частицы ($\pi$- и $K$-мезоны).

## 2.5. Проверка правила OZI

Правило OZI (Окубо, Цвейга, Иидзуки) [38] предсказывает сильное подавление процессов, описываемых диаграммами, в которых происходит образова-ние пары кварк-антикварк или ее аннигиляция в глюоны. Иными словами, в случаях, когда нарушается непрерывность кварковых линий (см. рис. 9).

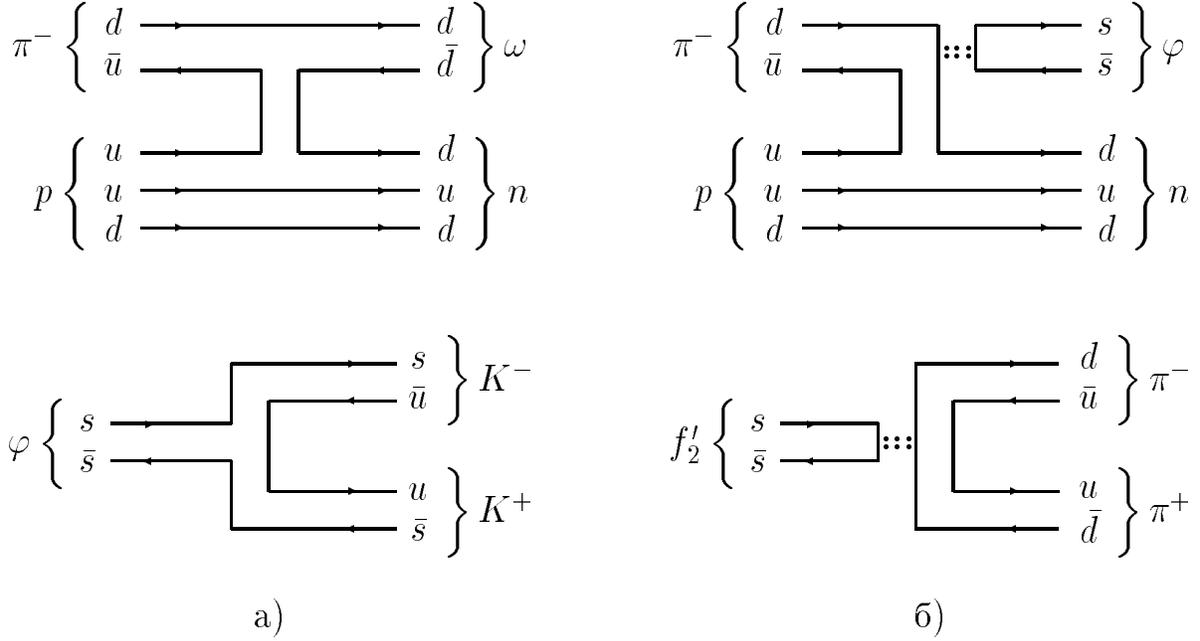

а)                                          б)

Рис. 9: Процессы: а) – разрешенные и б) – запрещенные правилом OZI.

Например, отношения квадратов матричных элементов распадов $\varphi$- и $\omega$-мезонов и $f_2'$- и $f$-мезонов, соответственно, на мезоны, не содержащие странные кварки, равны

$$|M(\varphi \to \pi^+\pi^-\pi^0)/M(\omega \to \pi^+\pi^-\pi^0)|^2 \approx 0.01,$$

$$|M(f_2' \to \pi^+\pi^-)/M(f \to \pi^+\pi^-)|^2 \approx 0.002.$$

В последнее время вопрос о тщательной проверке правила OZI приобрел особый интерес, что связано с поисками экзотических мезонных и барионных состояний. В этом случае в силу сложной цветовой структуры таких состояний можно ожидать существенного нарушения правила OZI при их образовании, что позволит легче выделить экзотические объекты на фоне подавленных правилом OZI состояний.

Следует заметить, что значительные нарушения правила OZI наблюдены в реакциях аннигиляции $p\bar{p}$ ($p\bar{p} \to \varphi\pi^0$, $p\bar{p} \to \omega\pi^0$ и др.

Выполнение правила OZI проанализировано сотрудничеством СФИНКС (ИТЭФ-ИФВЭ) в реакциях

$$\pi^- p \to \varphi + n \quad \text{и} \quad \pi^- p \to \omega + n$$

на установке LEPTON-F при импульсе $\pi^-$-мезонов 32.5 ГэВ/$c$ и в процессах образования систем $p\varphi$ и $p\omega$ в протон-ядерном взаимодействии на установке СФИНКС при энергии протонов 70 ГэВ [39]. В последнем случае изучались реакции $p + N \to (p\varphi) + N$ и $p + N \to (p\omega) + N$.

Данные по сечениям процессов образования $\varphi$- и $\omega$-мезонов использовались для проверки правила отбора OZI в адронных процессах. Анализ показал, что для пионных реакций отношение выходов этих мезонов $R(\varphi/\omega) \approx (3 \pm 1) \cdot 10^{-3}$ находится в хорошем согласии с предсказаниями наивной кварковой модели, основанными на данных об угле смешивания в нонете векторных мезонов и на правиле отбора OZI ($R(\varphi/\omega)|_{OZI} = \tan^2 \Delta\theta_V \approx 4 \cdot 10^{-3}$). Однако в протонных реакциях эффективное отношение выходов $\varphi$- и $\omega$-мезонов составляет $\approx (4 \pm 7) \cdot 10^{-2}$, т.е. имеет место сильное нарушение правила OZI в этих процессах. Возможно, это свидетельствует о существовании экзотической $s\bar{s}$-компоненты в кварковом составе протона.

Принимая во внимание то, что правило OZI имеет, по-видимому, более широкое применение, представляется важным его тщательная проверка в возможно более широком наборе процессов.

## 2.6. Поляризационные параметры в реакциях рождения странных частиц

Измерение спин-зависящих параметров рассеяния дополнительно к дифференциальным сечениям является важной экспериментальной задачей, решение которой облегчит изучение фундаментальных вопросов по барионной спектроскопии и динамике сильных взаимодействий с участием странных кварков. Знание поляризационных параметров необходимо для безмодельного восстановления амплитуд и осуществления однозначного парциально-волнового анализа. Несмотря на то, что поляризационное состояние $\Lambda$- и $\Sigma^{+}$- ($\Sigma^{-}$-) гиперонов полностью определяется по асимметрии их слабого распада, а для $\Sigma^{0}$ ($\Sigma^{0} \to \Lambda\gamma$) справедливо соотношение $\langle P_\Lambda \rangle = -1/3 P_\Sigma$, то есть нет необходимости во втором анализирующем рассеянии, экспериментальные данные в большинстве случаев недостаточны для прямого восстановления амплитуд. Реакции $\pi p \to K\Lambda(\Sigma)$ с экспериментальной точки зрения не относятся к числу простых. Нужно разделить процессы с рождением $\Lambda$ и $\Sigma^{0}$, которые являются фоновыми по отношению друг к другу. Разность масс этих гиперонов составляет 77 МэВ.

При $P_\pi < 3$ ГэВ/$c$ (резонансная область энергий, в которой обнаружено большое число нуклонных и изобарных состояний) желательны более подробные измерения в эксперименте с поляризованной протонной мишенью параметров поворота спина в $\pi^{+}p \to K^{+}\Sigma^{+}$ [40] для проведения нового ПВА, чтобы установить связь резонансов $\Delta^{++}$ с каналом $K^{+}\Sigma^{+}$. Отметим, что полный опыт в $\pi^{-}p \to K^{0}\Lambda$ и последующий ПВА [41] привели к обнаружению связи состояний $N(1650)S_{11}$, $N(1710)P_{11}$, $N(1720)P_{13}$ с $K^{0}\Lambda$. Является ли отсутствие аналогичных результатов по $\Delta$ следствием недостатка экспериментальных данных или имеется физическая причина существенного различия между $\Delta$ и $N^{*}$ — вопрос, ответ на который может быть получен на основе новых измерений.

При $P_\pi > 3$ ГэВ/$c$ измерение индивидуальных спиральных амплитуд в бинарных реакциях $\pi^- p \to K^0 \Lambda$, $\pi^- p \to K^0 \Sigma^0$, $\pi^+ p \to K^+ \Sigma^+$ обеспечит правильный выбор между конкурирующими феноменологическими моделями [42], основанными на фундаментально различающихся физических постулатах: редже-полюсные модели и модели сильного поглощения. И те и другие воспроизводят одинаково хорошо имеющиеся данные по дифференциальным сечениям и нормальной поляризации, но предсказывают резко различающееся поведение параметров поворота спина. Поэтому, начиная с конца 60-х годов, в работах, посвященных анализу появляющихся время от времени экспериментальных данных, неоднократно подчеркивается актуальность измерений параметров поворота спина. Поставленный с этой целью эксперимент S134 в ЦЕРНе в 1975 году закончился неудачно, так как требовались более совершенные трековые детекторы. Задача рассматривалась и в более широком плане — измерить параметры в кроссинг-симметричных реакциях с целью разделения вкладов векторного $K^*(890)$- и тензорного $K^*(1420)$-обменов с натуральной четностью [43]. При сравнении реакций с обменом гиперзаряда и реакций перезарядки ($\pi^- p \to \pi^0 n$ и $\pi^- p \to \eta n$ — соответственно $\rho$- и $a_2$-обмены) наблюдается аномальное различие в поведении дифференциальных сечений, которое может найти свое объяснение в рамках моделей, связанных с непертурбативной КХД [44].

Следующая интересная серия экспериментов относится к квазидвухчастичным реакциям $\pi^- p \to K^*(890)\Lambda$ и $\pi^- p \to K^*(890)\Sigma^0$ на водородной мишени. Доступно измерение элементов матрицы плотности $K^* \to K^+ \pi^-$, нормальной поляризации $\Lambda$, $\Sigma^0$ и угловых корреляций вторичных частиц в $K^*(890)\Lambda$, что позволяет получить информацию об элементах двойной матрицы плотности $K^*(890)\Lambda$. В результате удается восстановить шесть амплитуд $A_\pm, S_\pm, D_\pm$, сконструированных из спиральных амплитуд $F_{\lambda,\lambda'}^\mu$, где $\lambda$, $\lambda'$ и $\mu$ — спиральности протона, $\Lambda$ и $K^*$, соответственно, например, $A_\pm = (F_{++}^0 \mp i F_{+-}^0)/\sqrt{2}$. Могут быть восстановлены относительные фазы в пределах каждого из двух наборов амплитуд ($A^+, S^-, D^+$) и ($A^-, S^+, D^-$). Относительная фаза между этими двумя наборами может быть определена из эксперимента с продольно-поляризованной мишенью. В этих реакциях разрешен обмен с натуральной и ненатуральной четностью. Соответственно вклады обменов с натуральной и ненатуральной четностью в дифференциальное сечение определяются выражениями $(\rho_{11} + \rho_{1-1})d\sigma/dt$ и $(1 - \rho_{11} - \rho_{1-1})d\sigma/dt$, соответственно. Представляет интерес сравнение величин $(\rho_{11} + \rho_{1-1})d\sigma/dt$ для $\pi^- p \to K^*(890)\Lambda(\Sigma^0)$ и $d\sigma/dt$ для $\pi^- p \to K^0 \Lambda(\Sigma^0)$, а также соответствующих поляризаций. В данных, полученных на пузырьковых камерах, можно увидеть указание на существенное различие в поведении приведенных величин в области малых переданных импульсов. Требуются более точные экспериментальные данные, чтобы восстановить амплитуды, в частности, также, чтобы определить величину фазы между двумя наиболее важными амплитудами $A_+$ и $S_-$. Следует отметить, что спино-зависящие параметры реакции $\pi^- p \to K^*(890)\Lambda$, измеренные на водородной мишени, в той же степени информативны, что и поляризационные данные в $\pi^- p \to \rho^0 n$, полученные на поперечно-поляризованной мишени [45].

Другая задача, для решения которой желательны поляризационные измерения, относится к проблеме "запрещенного" рассеяния вперед в $\pi^- p \to K^+ \Sigma^-$, которое могло бы осуществляться благодаря обмену экзотическим $K\pi$-резонансом с $I = 3/2$. Если обмен не экзотический, а осуществляется двумя частицами (например, $K\pi$ или $\rho K^*$), то интерференция соответствующих амплитуд может привести к поляризации $\Lambda$ из распадов $\Sigma^-(1385) \to \Lambda\pi^-$. Пока пик при рассеянии вперед в этой реакции не удалось объяснить ни за счет каких-либо известных кинематических отражений, ни простыми $s$-канальными эффектами [45].

Существует по крайней мере два экспериментальных свидетельства того, что процессы с $\Lambda$ и $\Sigma^0$ могут быть разделены. В эксперименте с магнитным спектрометром это удалось сделать за счет хорошего разрешения по недостающей массе к $K^0$ при 4 ГэВ/$c$ [46]. В водородных камерах измерялись кинематические параметры всех заряженных частиц, и разделение осуществлялось по $\chi^2$. Таким образом, если в магнитном спектрометре вершинный детектор будет обладать свойствами пузырьковой камеры, близким к $4\pi$ аксептансом, и высоким импульсным разрешением, то перечисленные выше реакции будут изучены на большом статистическом материале. Таким вершинным детектором может быть время-проекционная дрейфовая камера.

## 2.7. Поиск аномальных явлений в пороговом поведении сечения рождения двух гиперонов в $pp$-взаимодействии

С целью проверки теоретических модельных предсказаний о существовании гиперон-гиперонных связанных состояний предлагается провести исследование порогового поведения реакций с рождением пар гиперонов или каскадных гиперонов, сопровождающихся вылетом двух $K^+$-мезонов и поиск адронных связанных состояний с двойной странностью.

Исследование порогового поведения реакций с рождением дигиперона или каскадных гиперонов, сопровождающихся вылетом двух $K^+$-мезонов, в протон-протонных и протон-ядерных столкновениях представляет большой теоретический и экспериментальный интерес. Связанные состояния со странностью $-2$ в системе двух гиперонов и гиперон-нуклона предсказаны КХД, но до сих пор есть лишь указания на гиперон-нуклонные квазиядра.

Гиперон-гиперонные связанные состояния пока не наблюдались, хотя именно здесь могут быть найдены пути к решению проблемы существования "странной адронной материи". Опыт предполагается провести на внутреннем пучке синхротрона ИТЭФ. В качестве мишени будут использованы тонкие ядерные мишени и криогенная мишень-инжектор замороженных микрокапель различных газов (pellet target), в том числе водорода и дейтерия. Будут изучаться следующие реакции:

| Реакция | | | | $P_{\text{порог}}$, ГэВ/$c$ |
|---|---|---|---|---|
| $pp$ | $\rightarrow$ | $H(2220)$ | $+\quad K^+K^+$ | 4.45 |
| $pp$ | $\rightarrow$ | $\Lambda\Lambda$ | $+\quad K^+K^+$ | 4.485 |
| $pp$ | $\rightarrow$ | $\Lambda\Sigma^0$ | $+\quad K^+K^+$ | 4.757 |
| $pp$ | $\rightarrow$ | $\Sigma^+\Sigma^-$ | $+\quad K^+K^+$ | 5.08 |
| $pp$ | $\rightarrow$ | $\Xi^0 n$ | $+\quad K^+K^+$ | 4.567 |
| $pp$ | $\rightarrow$ | $\Xi^- p$ | $+\quad K^+K^+$ | 4.585 |
| $pp$ | $\rightarrow$ | $ppK^-K^-$ | $+\quad K^+K^+$ | 6.9 |

В предлагаемой постановке опыта ускоритель является частью экспериментальной установки. $K^+$-мезоны будут выделяться многослойным детектором-идентификатором, а их треки регистрироваться пропорциональными камерами.

Реакции $pp$-взаимодействия с образованием двух $K^+$-мезонов имеют весьма небольшие сечения и, поэтому, исследованы недостаточно подробно, хотя некоторые их аспекты представляют несомненный интерес. Прогресс в технике эксперимента позволяет надеяться получить новые сведения об этих процессах.

Приведем некоторые из известных данных. В работе [47] при импульсе налетающих протонов 5.4 и 5.9 ГэВ/$c$ приведены значения $23.4 \pm 10.0$ и $360 \pm 90$ нб для инклюзивного сечения рождения двух $K^+$-мезонов. Холмгрен и др. [48] получили при импульсе протонов 10 ГэВ/$c$ величину сечения $7 \pm 5$ мкб для канала $KK\Xi^-$. В [49] измерено инклюзивное сечение рождения $\Xi^-$ при 19.1 ГэВ/$c$: $30 \pm 20$ мкб.

Грубые оценки сечений реакций с образованием двух $K^+$-мезонов вблизи порога рождения гиперонных пар приводят к величинам этих сечений от 100 пб до 1 нб.

Реакция $pp$-взаимодействия, когда в конечном состоянии образуются два $K^+$-мезона, интересна еще и потому, что в ней должна проявиться гипотетическая частица — $H^0$-дибарион со странностью $-2$, предсказанная Джаффе [50] в 1977 году. Работа [47] как раз и посвящена поиску этого дибариона. В ней получены верхние пределы сечений рождения $H^0$. Так, при 5.1 ГэВ/$c$ и для интервала масс $2.1 \div 2.33$ ГэВ получено ограничение – 50 нб. Для импульса протонов 5.4 ГэВ/$c$ в интервале $2.1 \div 2.23$ ГэВ верхний предел – 40 нб, а для интервала $2.23 \div 2.35$ ГэВ — 30 нб.

В работах [51] и [52] в фотоэмульсии наблюдались гиперядра $_{\Lambda\Lambda}Be^{10}$ и $_{\Lambda\Lambda}He^6$, что делает область $1.9 \div 2.2$ ГэВ для массы $H^0$ маловероятной. Однако, в последнее время были предприняты попытки найти этот дибарион со временем жизни 10 нсек и выше (область масс 2.15 ГэВ и ниже). Так, в работе [53] получено ограничение на сечение рождения дибариона со временем жизни больше $10^{-8}$ сек в столкновениях протонов с импульсом 24.1 ГэВ/$c$ с ядрами платины — 1 мб/ср.

В Брукхейвене продолжаются 4 эксперимента по поиску $H$-дибариона. Приводимые ниже сведения взяты из небольшого обзора [54].

В эксперименте E813 исследуется реакция $K^-p \to K^+\Xi^-$ с остановкой гиперона в дейтериевой мишени и образованием квазиатома, который мог бы превратится в $H^0$ и нейтрон. Снимается спектр нейтронов, в котором искомой реакции должна соответствовать монохроматическая линия. Сигнал о существовании $H^0$ пока не обнаружен, опыт продолжается.

Эксперимент E888 нацелен на регистрацию распадов $H^0$ на лету по модам $H^0 \to \Lambda n$ или $\Sigma^0 n$. Дибарионы рождаются пучком протонов с импульсом 24 ГэВ/$c$ на ядерной мишени. Результат пока отрицательный, опыт продолжается.

Эксперимент E836 — это развитие E813 с использованием одной мишени — $^3He$. В этом случае $K^- +^3He \to K^+H^0n$. Экспериментальных данных пока нет.

В работе [55] при столкновении релятивистских ионов $Au$ с ядрами платины получено ограничение на сечение рождения $H^0$-ядер ($H^0\,^3He$) и ($H^0\,d$) порядка $10^{-5}$ мб/ГэВ$^2$ около $y = 0.6$ и $p_t/Z = 0.18$ ГэВ/$c$.

В ежегодном отчете института ядерных исследований университета Токио приведена работа [56], где исследуется рождение $H^0$ на ядре углерода $K^-$-мезонами, и для области масс $2200 \div 2230$ МэВ получено ограничение $0.6 \div 0.7\%$ от сечения рождения $\Xi^-$.

В работах [57] и более ранних тех же авторов обнаружена целая группа дибарионов, но ограниченность статистики не позволяет считать их надежно установленными. Сечения рождения $H^0$ протонами с импульсом 10 ГэВ/$c$ на пропане, согласно этим работам, равно 40 нб ($m_H = 2.174$ ГэВ). Сечения рождения состояний со странностью $-2$ и массой, большей, чем масса двух $\Lambda$, равны десяткам микробарн.

В работе [58] при исследовании взаимодействий нейтронов с ядрами обнаружено два события, допускающих интерпретацию как рождение и последующий распад $H^0$. Величина сечения оценена авторами в 138 нб на ядро углерода. Нейтроны имели широкий спектр с максимумом около 8 ГэВ/$c$.

Изучение подпорогового образования состояний с удвоенной странностью проводилось в ИТЭФ в исследовании аннигиляций антипротонов с ядрами ксенона. Исследовалась реакция $p + Xe \to K^+K^+YY + X$ и получено лучшее в настоящее время ограничение на вероятность рождения $H$-дибариона ($S = -2$) в антипротонных взаимодействиях [59, 60].

Таким образом видно, что исследование взаимодействий гиперонов и, особенно, поиски $H$-дибариона ($S = -2$) являются в настоящее время одним из важнейших направлений в физике элементарных частиц. Такие исследования проводятся почти во всех крупнейших ускорительных лабораториях мира. Изучение поведения гиперонных пар на пороге рождения в $pp$-взаимодействиях, несомненно, даст важнейшую информацию как для проверки модельных предсказаний и анализа взаимодействий гиперонов, так и решения проблемы "странной материи".

## 2.8. Поиск распада $\omega \to \pi^+\pi^-\gamma$

Исследования электромагнитных распадов адронов играют важную роль в физике элементарных частиц. Эти процессы, определяемые взаимодействием реальных и виртуальных фотонов с электрическими зарядами кварков, позволяют получить уникальную информацию о характере различных кварковых конфигураций в адронах. Большой интерес вызывают радиационные распады легких векторных мезонов. Для их анализа была развита теория эффективного кирального лагранжиана, описывающего взаимодействие векторных и псевдоскалярных мезонных полей. В рамках этого подхода удалось описать распады типа $V \to P\gamma$ (например, $\omega, \rho \to \pi\gamma$ и т.п.), введя всего одну константу связи векторных и псевдоскалярных полей $g_{VVP} = 7 \div 10$ ГэВ$^{-1}$. Описание же распадов $V \to PP\gamma$ ($\omega, \rho, \phi \to \pi\pi\gamma$) наталкивается на неопределенности, связанные с вкладом т.н. аномального Весс-Зуминовского члена в лагранжиане, а также на возможный вклад четырехкварковых скалярных мезонов $V \to S\gamma \to \pi\pi\gamma$. Поэтому измерение распадов $V \to PP\gamma$ позволило бы проверить на новом классе распадов теорию эффективного кирального лагранжиана, более точно определить параметры модели и оценить роль многокварковых состояний в этих распадах.

Однако до настоящего времени ни один из таких распадов не был найден в эксперименте (исключение составляет распад $\rho \to \pi^+\pi^-\gamma$, который идет за счет тормозного излучения, что не позволяет наблюдать обсуждаемые здесь эффекты, имеющие существенно меньшую вероятность). Наиболее близко существующие эксперименты приблизились к обнаружению распада $\omega \to \pi^+\pi^-\gamma$, парциальная вероятность которого ожидается на уровне $(3 \div 5) \cdot 10^{-4}$, при верхней границе в $3 \cdot 10^{-3}$ (на 95% уровне достоверности), достигнутой в эксперименте ASTERIX в CERN [61]. Распад $\omega \to \pi^+\pi^-\gamma$ имеет еще и особый интерес. В нем вероятности тормозного механизма и структурного (Весс-Зуминовского) имеют один и тот же порядок, поэтому, в принципе, можно наблюдать интерференционные эффекты этих механизмов. Распад $\omega \to \pi^+\pi^-\gamma$ искался с 60-х годов, сначала в пузырьковой камере, а затем на более сложных установках: на встречных $e^+e^-$-пучках в Новосибирске, на больших магнитных спектрометрах ЛЕПТОН-Ф на пучке 38 ГэВ/$c$ отрицательных пионов в ИФВЭ и на ASTERIX на пучке останавливающихся антипротонов в CERN.

Ускоритель ИТЭФ имеет пучки пионов и протонов, оптимальные для поиска этого распада. Наибольшее сечение рождения $\omega$ имеет в реакции $\pi^-p \to \omega n$ при энергиях вблизи порога (порог 1.1 ГэВ/$c$). Однако, для поиска распада требуется высокоточная регистрация импульсов нейтрона и $\gamma$-кванта. Другой путь — это рождение $\omega$ в реакции $pp \to pp\omega$. Сечение этой реакции вблизи порога недавно было измерено на ускорителе Сатурн в Сакле. Оно составляет 1 мкб при энергии протонов $T_p = 1940$ МэВ (порог 1890 МэВ, $p = 2.67$ ГэВ/$c$). При интенсивности протонов $10^7$ за сброс, 1000 сбросах в час и длине жидководородной мишени 25 см будет рождаться $10^4$ $\omega$/час. Если ориентироваться на предсказываемую вероятность распада $5 \cdot 10^{-4}$ и эффективность регистрации распада $0.1 \div 0.01$, то за 100 часов работы можно будет зарегистрировать $50 \div 5$ распадов. Таким образом имеются высокие шансы обнаружить этот распад на ожидаемом уровне.

## 2.9. Измерение электромагнитных форм-факторов $\eta$-, $\eta'$- и $\omega$-мезонов

Структура адронов является важным тестом для непертурбативного режима квантовой хромодинамики. Распределение заряда и магнетизма в адронах описывается соответствующими форм-факторами. В то время как пространственно-подобные форм-факторы изучены довольно хорошо, например, в рассеянии электронов, во времениподобной области электромагнитные форм-факторы известны плохо. Они могут быть изучены путем наблюдения излучения дилептонов или Далитц-распадов барионных резонансов или нейтральных мезонов.

Изучение переходных процессов типа рис. 10 во времениподобной обла-

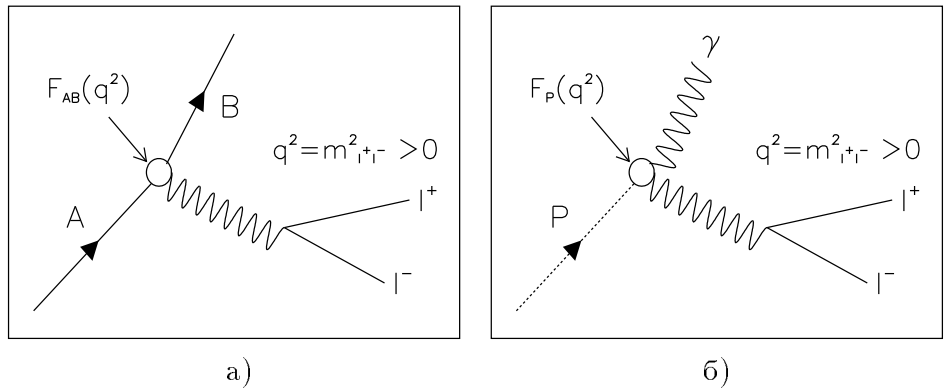

Рис. 10: Определение переходного форм-фактора $F_{AB}$ (а) и переходный форм-фактор при Далитц-распаде псевдоскалярного мезона (б).

сти $(q^2 = m_{ll}^2 > 0)$ для нейтральных мезонов позволяет определить переходный форм-фактор, характеризующий сложную динамическую структуру переходной вершины. Это возможно осуществить путем измерения спектра эффективных масс лептонных пар

$$d\Gamma/dq^2 = (d\Gamma/dq^2)_{\text{ТОЧЕЧНОЕ}} \cdot |F_{AB}(q^2)|^2 \qquad (q^2 = m_{l^+l^-}^2),$$

где $(d\Gamma/dq^2)_{\text{ТОЧЕЧНОЕ}}$ – спектр масс дилептонов при точечном взаимодействии по КЭД.

В работе [62] на установке "Lepton-G", облученной пучком $\pi^-$-мезонов с импульсом 32.5 ГэВ/$c$ Серпуховского ускорителя, изучались переходные процессы

$$\eta \rightarrow \mu^+ \mu^- \gamma ,$$

$$\eta' \rightarrow \mu^+\mu^-\gamma \,,$$
$$\omega \rightarrow \pi^0\mu^+\mu^-$$
$$\searrow 2\gamma$$

в реакциях

$$\pi^-p \rightarrow \mu^+\mu^-\gamma n \,,$$
$$\pi^-p \rightarrow \mu^+\mu^-\gamma\gamma n \,.$$

Как видно из рис. 11, данные по переходному форм-фактору $\eta$-мезона имеют хорошую статистическую обеспеченность (образец содержал $\approx 500$ событий распадов $\eta \rightarrow \mu^+\mu^-\gamma$). Однако данные по $\eta'$- и $\omega$-мезонам статистически не

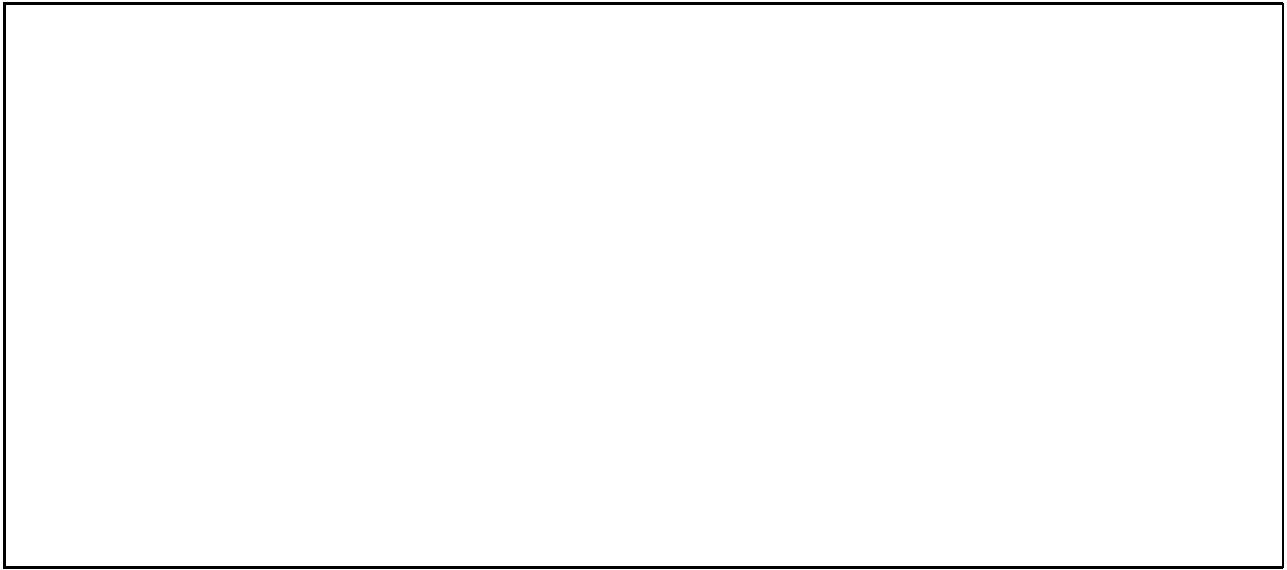

а)                          б)                          в)

Рис. 11: Экспериментальные данные по переходным форм-факторам и предсказания теоретических моделей. Пунктирная кривая на рис. б) – предсказания нелокальной кварковой модели. На рис. в): кривая 1 – результат экстраполяции экспериментальных данных, кривая 2 – предсказания VDM с модифицированным $\rho$-пропагатором, кривая 3 – предсказания VDM.

обеспечены (образец $\eta'$ содержал $\sim 25$ событий, а $\omega$-мезонов – $\sim 50$). Сравнение экспериментальных данных с расчетами по модели векторной доминантности (VDM) позволяет сделать следующее заключение.

1. Для $\eta$-мезона экспериментальные данные удовлетворительно согласуются с предсказаниями модели VDM.

2. Для $\eta'$-мезона экспериментальные данные имеют слишком большие ошибки, не позволяющие провести надежное сравнение с расчетами по модели VDM.

3. Экспериментальные данные для $\omega$-мезона противоречат предсказаниям VDM, особенно при больших $q^2 = m^2_{\mu^+\mu^-}$. Разница превышает 4 стандартных отклонения.

Учитывая, что приведенные данные получены только в одной работе, представляется важным провести тщательные измерения переходных формфакторов нейтральных мезонов $\eta$, $\eta'$, $\omega$, а также $\varphi$-мезона ($\varphi \to \eta l^+ l^-$).

При импульсе $\pi^-$-мезонов $4 \div 5$ ГэВ/$c$ сечение образования нейтральных мезонов намного выше, чем при импульсе 32.5 ГэВ/$c$. Например, сечение реакции $\pi^- p \to n\omega$ при импульсе $3 \div 5$ ГэВ/$c$ составляет $\approx 1.5 \div 2.0$ мб, а при импульсе 32.5 ГэВ/$c$ всего лишь $\approx 0.003$ мб. Аналогичная картина наблюдается и для $\eta$-, $\eta'$-, $\varphi$-мезонов. Подобные измерения на ускорителе ИТЭФ вполне возможны при статистике, по крайней мере, на порядок превышающей использованную в работе [62] на Серпуховском ускорителе. Изменение импульса начальных $\pi^-$-мезонов позволит образовать вторичные мезоны в покое, что представляет интерес для исследования их поведения.

Важным является также измерение прямых радиационных распадов векторных и псевдоскалярных мезонов типа

$$V \to P + \gamma\,,$$
$$P \to V + \gamma\,,$$

например, $\omega \to \pi^0\gamma$, $\omega \to \eta\gamma$, $\eta' \to \omega\gamma$ и др.

Точное определение их радиационных ширин и форм-факторов (также как и переходных процессов) позволит провести надежное сравнение с предсказаниями модели VDM и других, наблюдать возможную аномалию в магнитных моментах кварков, получить другие важные сведения.

Здесь следует упомянуть также о важности широкой программы изучения электромагнитных форм-факторов барионов таких, например, как изображенный на рис. 12.

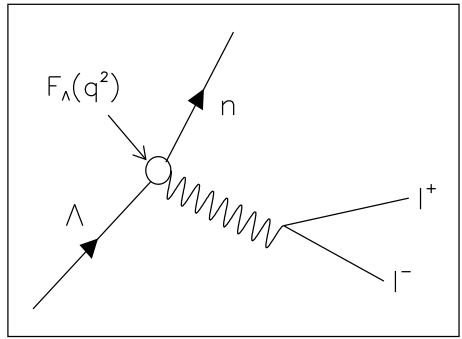

Рис. 12: Переходный форм-фактор $F_{\Lambda n}$.

# 3. Адрон-ядерные процессы

## 3.1. Измерение размеров области образования протонов в глубоконеупругих ядерных реакциях с целью оценки размера флуктона

Существует класс впервые выделенных в ИТЭФ [63] т.н. глубоконеупругих ядерных реакций (ГНЯР), которые протекают не на отдельных нуклонах ядра, а на тесных группах нуклонов — флуктонах. Хорошим примером ГНЯР являются реакции с образованием кумулятивных частиц, например, протонов, летящих назад в лаб. системе. Однако, в результате ГНЯР могут образоваться протоны (и их экспоненциально больше) которые летят вперед. На тяжелых ядрах доля таких протонов по сравнению с квазисвободными доминирует.

С современной точки зрения флуктон представляет собой многокварковое образование в ядре или, может быть, капельку барионно-насыщенной кварк-глюонной плазмы (КГП). Последнее обстоятельство привлекает особое внимание к определению плотности флуктонов, а следовательно — к оценке их размеров.

ГНЯР подробно изучались в ИТЭФ в течение многих лет [64, 65]. Их нетривиальное свойство (скейлинг, аномальная $A$-зависимость, характер выхода на скейлинг и т.д.) привели к следующей модели протекания ГНЯР. Налетающая частица двигается сквозь ядро, последовательно возбуждая на своем пути флуктоны и теряя при этом приблизительно 1 ГэВ энергии на 1 ферми пути [66]. Изучение корреляций позволили методом Копылова и Подгорецкого [67] не только измерить длину пути, но и установить факт того, что поперечный размер области, из которой вылетает нуклон, в несколько раз меньше продольного размера [68]. Оценка поперечного размера дает величину $1 \pm 1$ ферми, т.е. известна сегодня со 100%-ой ошибкой. Было найдено, что продольный размер зависит от угла наблюдения, поскольку летящая вторичная частица догоняет или, напротив, улетает от налетающей частицы. Наконец, было показано [69], что продольный размер уменьшается для частиц, образующихся под действием частиц с относительно малой начальной энергией, в соответствии с указанными выше потерями ими энергии. Продольный размер трубки вдоль траектории частицы, как и следовало ожидать в рамках вышеописанной модели, порядка размера ядра мишени. Поперечный размер трубки зависит от размеров налетающей частицы, размера флуктона и, может быть, от рассеяния в ядерном веществе вторичных частиц. Так или иначе, если известны поперечные размеры, то известно ограничение на размер флуктона сверху. Мы знаем из значения кумулятивного числа минимально возможную массу флуктона, необходимую для образования нуклона данного импульса, т.е. можно оценить минимальную плотность флуктона и близость этого объекта к границе кварк-глюонной плазмы на известной диаграмме плотность–температура для ядерного ве-щества.

Предлагается измерить корреляционные функции двух протонов с малыми относительными импульсами, которые образуются в реакции

$$p + A(C, Pb) \rightarrow pp + X \;.$$

Протоны должны регистрироваться в импульсном диапазоне $500 \div 800$ МэВ/$c$. Начальная энергия протонов в диапазоне от 2 до 10 ГэВ. Можно начать с энергии порядка 2 ГэВ, что достаточно для возбуждения одного флуктона, но ограничивает возбуждение нескольких флуктонов вдоль трубки, делая продольный размер малым. Это обстоятельство не позволяет увидеть большой разницы в продольном и поперечном размере, как это было в [68], но позволяет найти поперечный размер стандартным способом по величине пика, который, в основном, обусловлен сильным взаимодействием. Полная программа исследований должна включать измерение пар протонов, вылетающих на углы $5^o$, $90^o$, $175^o$. Это даст возможность оценить влияние рассеяния, изучить изменение размеров с изменением порядка кумулятивности. Измерение можно начать с угла порядка $5^o$, тогда предложенный эксперимент полностью совместим с предложением изучения фазовой плотности частиц. Предполагаемая точность измерений как перпендикулярного, так и продольного размеров 0.2 ферми.

## 3.2. Исследование взаимодействия флуктонов (многокварковых мешков) в ядро-ядерных столкновениях

Обычно подразумевается, что столкновение быстрого нуклона с ядром сводится к столкновениям с отдельными квазисвободными нуклонами ядра. Однако, в случае ГНЯР происходит взаимодействие налетающей частицы с плотной флуктуацией ядерной материи — флуктоном ($Фл.$), который можно представлять себе как многокварковый мешок и, может быть, капельку кварк-глюонной плазмы. ГНЯР были выделены на фоне квазисвободных взаимодействий с отдельными нуклонами путем проведения измерений в кинематически запрещенной для $NN$-взаимодействий области. Например, протоны могут лететь назад в л.с., что запрещено в $NN$-столкновениях. ГНЯР обладают, как оказалось, рядом удивительных нетривиальных свойств, неизвестных в $NN$-взаимодействии.

В ядро-ядерных столкновениях, конечно, есть как $NN$-, так и $N\textit{Фл.}$-процессы, которые могут быть изучены при $NA$-взаимодействиях, но есть $\textit{Фл.}\textit{Фл.}$-процессы, не сводящиеся к $NN$- и $NA$-взаимодействиям.

Можно ли их выделить? Оказывается, можно кинематически [70]. На рис. 13 под кривой 2.2 показана такая кинематическая область при быстротах $y \sim 0$ при больших $p_T$, которая соответствует $\textit{Фл.}\textit{Фл.}$-взаимодействиям. Регистрируя события в этой области, измеряя спектры частиц в этой области, изучая их $A$-зависимость, можно:

- открыть новое явление — дважды кумулятивное образование частиц;

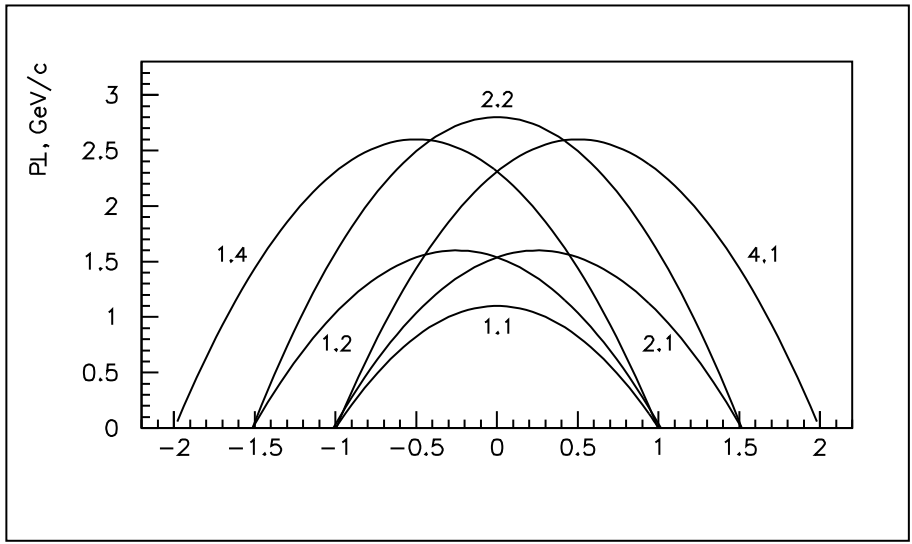

Рис. 13: Различные кинематические области в плоскости $p_T(y)$, ограниченные сверху кривыми 1.1, 1.2 и 1.4, 2.1 и 4.1, 2.2, которые соответствуют $NN$-, $N\Phi_\text{л}$.-, $\Phi_\text{л}N$-, $\Phi_\text{л}\Phi_\text{л}$.-соударениям.

- изучить свойства $\Phi_\text{л}\Phi_\text{л}$.-взаимодействия, которое может оказаться нетривиальным;

- изучить более плотные, чем флуктоны, образования, т.е. продвинуться по шкале $\rho/\rho_0$ в плоскости $(T, \rho/\rho_0)$ для поисков кварк-глюонной плазмы.

### 3.3. Исследование свойств ядерной материи на малых межнуклонных расстояниях в подпороговом образовании адронов

Исследование свойств горячей и плотной адронной материи, образующейся в столкновениях релятивистских ядер, является одной из самых актуальных задач современной ядерной физики. Интерес к этой области связан с поисками нового состояния адронной материи — кварк-глюонной плазмы [71] и предсказанных мотивированными КХД моделями эффектов восстановления киральной симметрии [72].

Изучение ядро-ядерных взаимодействий проводится сейчас в широком диапазоне энергий столкновений от $1 \div 2$ ГэВ/нуклон (SIS GSI) и 10 ГэВ/нуклон (AGS BNL) до 200 ГэВ/нуклон (SPS CERN). В начале следующего тысячелетия начнут работать ядерные коллайдеры с энергией несколько ТэВ/нуклон (RHIC, LHC).

Поскольку сложность интерпретации результатов экспериментов существенно возрастает с увеличением энергии и размеров сталкивающихся ядер, эксперименты на пучках протонов и легких ядер при энергиях несколько ГэВ/нуклон являются необходимым этапом изучения столкновений высокоэнергичных тяжелых ядер.

Изучение структуры ядра на малых межнуклонных расстояниях является одним из актуальных направлений релятивистской ядерной физики. Особую роль в этих исследованиях играют эксперименты по подпороговому рождению адронов в протон-ядерных взаимодействиях. На ускорителе ИТЭФ были выполнены эксперименты по подпороговому образованию антипротонов [73] и $K^+$-мезонов [74]. Специальный выбор кинематических условий измерений позволил выделить область доминирования прямого механизма рождения и впервые в сильных взаимодействиях получить информацию об импульсном распределении внутриядерных нуклонов на межнуклонных расстояниях меньших, чем доступные сегодня для изучения в экспериментах на электронных пучках. На рис. 14 сплошной кривой представлено импульсное распределение нуклонов в ядре $Be$, извлеченное из энергетической зависимости подпорогового образования $K^+$-мезонов. Черными квадратами показаны существующие сегодня данные, полученные из анализа $(e, e'p)$-реакций.

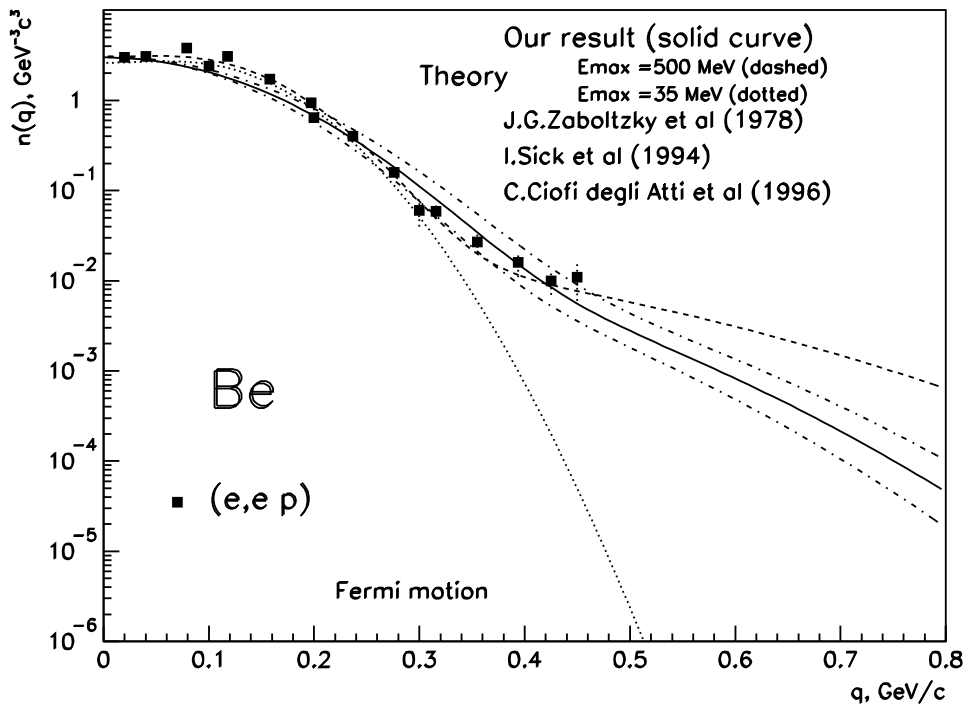

Рис. 14: Импульсное распределение нуклонов в ядре $Be$

В настоящее время в ИТЭФ проводится исследование процесса рождения подпороговых $K^-$-мезонов протонами на ядрах. Кроме самостоятельного интереса, связанного с полным отсутствием данных об этих реакциях, эти эксперименты интересны еще и тем, что содержат информацию о предсказанных киральными моделями эффектах модификации масс каонов в барионном окружении. Теоретические расчеты (смотри, например, [72]) свидетельствуют, что, в отличие от положительно заряженного каона, масса отрицательного каона в ядре меньше его массы в вакууме на 20% уже при нормальной ядерной плотности. Эксперимент еще не закончен, однако проведенный совместный анализ полученных данных об образовании $K^+$- и $K^-$-мезонов свидетельствует о наличии такого эффекта.

Согласно киральным моделям, основанным на использовании эффективных лагранжианов, не только свойства мезонов, но также свойства нуклонов и гиперонов заметно изменяются в среде уже при нормальной ядерной плотности [75]. Именно подпороговые реакции могут стать чувствительным инструментом для исследования модификации масс, так как из-за дефицита энергии столкновения уменьшение массы рожденной системы частиц всего на $10 \div 20$ МэВ приводит к увеличению в несколько раз расчетной вероятности ее образования [76].

Предлагаемая программа исследования ядро-ядерных столкновений относится к неизученной и недоступной для исследования на других ускорителях области энергий ионных пучков $2 \div 4.5$ ГэВ/нуклон. Эта область соответствует подпороговому и околопороговому образованию мезонов и барионов, где из-за дефицита энергии столкновения число открытых каналов их генерации минимально. Это обстоятельство существенно для однозначной интерпретации результатов измерений. Имеющиеся сегодня экспериментальные и расчетные данные [77] дают возможность провести реалистические оценки скорости набора информации.

Исследование рождения антибарионов, барионных резонансов, скалярных и векторных мезонов в столкновениях легких ядер дейтерия и гелия с легкими мишенями представляется чрезвычайно важным и информативным этапом общей программы, необходимым для однозначной интерпретации результатов экспериментов с тяжелыми ионами. Аномально большие сечения подпорогового рождения антипротонов по сравнению с наблюдаемыми в протон-ядерных реакциях были обнаружены в LBL [78] и GSI [79] в столкновениях средних и тяжелых ядер. Это явление удалось описать включением каналов генерации подпороговых антипротонов в каскадных процессах с участием изобар [80], либо уменьшением массы рожденных антипротонов в барионном окружении [81]. Оба объяснения требуют увеличения барионной плотности в зоне перекрытия сталкивающихся ядер. Однако недавние эксперименты в КЕК [82] показали, что сильный эффект 100(!) наблюдается уже в реакциях, индуцированных дейтронами, где трудно ожидать существенного возрастания плотности. Эффект увеличения выхода антипротонов при облучении легких мишеней пучком $^4He$ составляет 1000. Сравнение фрагментарных данных [82] по подпороговому рождению $K^-$-мезонов в дейтрон-ядерных взаимодействиях с полученными в ИТЭФ в протон-ядерных столкновениях демонстрируют наличие столь же сильного эффекта для отрицательных каонов. Эти наблюдения указывают на то, что природа этого неожиданного и, по-видимому, универсального для всех рождающихся систем явления может быть изучена в экспериментах с легкими ядрами. Сегодня информация о подпороговом и околопороговом образовании легкими ядрами других скалярных мезонов, всех векторных мезонов и барионных резонансов полностью отсутствует. Данные по рождению мезонных и барионных резонансов легчайшими ядрами важны как для исследования динамики развития явлений, зависящих от температуры и плотности ядерного вещества, так и для изучения эффектов изменения их свойств в барионном окружении.

Поиски предсказанных киральными моделями эффектов модификации свойств адронов в плотной барионной среде проводятся сейчас во многих научных центрах. Экспериментальные данные по образованию мюонных пар в столкновениях ядер $S+Au$ и $S+W$ при энергии 200 ГэВ/нуклон, полученные в CERN коллаборациями CERES [83] и HELIOS-3 [84], интерпретируются как свидетельство уменьшения массы $\rho$-мезона [85, 86]. Коллаборация KAOS (GSI) сообщила о наблюдении существенного уменьшения массы $K^-$-мезона в столкновениях ядер $Ni + Ni$ при энергии 2 ГэВ/нуклон [87]. Предсказанная достаточно большая величина эффектов изменения свойств адронов стимулировала их поиски при нормальной ядерной плотности. Коллаборация TAGX (Токио) зарегистрировала значимое уменьшение массы $\rho$-мезонов в процессе их фоторождения на легких ядрах в околопороговой области энергии [88]. Выполняемый сейчас эксперимент E-325 в KEK направлен на поиски модификации свойств $\phi$-мезона в протон-ядерных взаимодействиях при 12 ГэВ [89].

В случае столкновения легких ядерных систем кинематические характеристики продуктов распада рождающихся объектов минимально искажены взаимодействиями в конечном состоянии, что предоставляет возможность изучать их свойства, используя адронные распады, ширины которых превосходят лептонные на несколько порядков величины [90]. О возможности получения обширной информации уже в соударениях легких ядер свидетельствует наблюденное в KEK [82] рождение антипротонов с импульсами до 1.5 ГэВ/$c$ в соударениях ядер гелия-4 с энергией 2.5 ГэВ/нуклон с углеродом. Инвариантная масса рожденной системы при этом превосходит 2 ГэВ, а наблюдаемое сечение составляет 30 мкб.

Уже полученные в KEK, GSI и ИТЭФ данные по подпороговому рождению каонов и антипротонов в протон-ядерных столкновениях будут служить основой для выделения специфических ядро-ядерных эффектов.

## 3.4.  Исследование явления ядерной критической опалесценции

При увеличении плотности ядерной материи в несколько раз взаимодействие пиона со средой становится настолько сильным, что он превращается из частицы в квазичастицу, которую А.Б. Мигдал называл "0-звуковым возбуждением" ядерного вещества [91]. При достаточной плотности ядерного вещества полная энергия пиона становится равной нулю, и возникают условия для развития Бозе-конденсации таких квазичастиц (пионный конденсат). Максимальная энергия связи 0-звуковых пионов определяет их критический импульс, который, согласно расчетам, близок к импульсу Ферми ядра ($k_C \approx p_F$). Направление волнового вектора $k_C$ 0-звуковой стоячей волны в ядре определяет направление, в котором мог бы происходить фазовый переход ядерной среды из жидкой фазы в фазу жидкокристаллическую, то есть в фазу, где периодическая структура распределения нуклонов возникает в одном измерении, а не в трех измерениях, как в кристалле. Гипотеза, что такой фазовый переход мог иметь место в нормальных ядрах, была отвергнута экспериментально [91], однако в конечных ядрах фазовый переход, если бы он даже и существовал, не мог бы приближаться к поверхности ядра на расстояние ближе, чем $\Delta = h/k_C$. Таким образом, даже и в случае развитого фазового перехода жидкокристаллическая фаза была бы скрыта под жидким ядерным слоем глубиной примерно в 1 ферми.

В работе М. Эриксон и Дж. Делорма [92] было показано, что вследствие сильных флуктуаций ядерной плотности оказывается возможным локальный фазовый переход, подобный явлению критической опалесценции жидкостей: из-за флуктуаций возникает множество короткоживущих микропузырьков новой фазы, делающих жидкость полупрозрачной. Предкритические ядерные эффекты обнаружить и закрыть очень трудно, поскольку они не только редки, как всякий флуктуационный процесс, но и скрыты жидкой фазой поверхности ядра. Однако, как было отмечено в [92], предкритические эффекты имеют и свое преимущество. Если частица налетает на ядро, в котором фазовый переход уже произошел, то угол между направлением падения и осью жидкого кристалла может быть произвольным. В случае же эффекта критической опалесценции локальная кристаллизация (возникновение тонких плоских слоев протонов и нейтронов) происходит по направлению налетающей частицы. Поскольку импульс налетающей частицы порядка 1 ГэВ/$c$, продольный размер флуктуации должен быть порядка нескольких десятых ферми. В случае развитого фазового перехода из-за того, что направление 0-звуковой волны неизвестно, невозможно вычислить полную энергию взаимодействия налетающей частицы с 0-звуковым пионом, тогда как в случае предкритического явления, когда ось флуктуации совпадает с направлением налетающей частицы, вычисление полной энергии взаимодействия возможно: $m_r^2 = -k^2 + 2kp_{in} + m_{in}^2$, где $p_{in}$ и $m_{in}$ — импульс и масса налетающей частицы, $k$ — импульс опалесцентного пиона, а $m_r$ - масса образованного резонанса.

Опалесцентный мезон можно представить как $t$-канальный пион, поглощаемый ядерной материей с образованием пары квазичастиц нуклон-дырка (рис. 15$a$). Рассмотрим взаимодействие мягкого нуклона с опалесцентным пионом (рис. 15$c$). Протон с импульсом 125 МэВ/$c$ ($T_p = 8$ МэВ) резонансно упруго отражается назад, поглощая опалесцентный пион ($m_r = m_{in} = m_N$). Поскольку в стоячей волне опалесцентные мезоны возникают парами с импульсами, направленными в противоположные стороны, протон, отраженный от одного опалесцентного пиона, может резонансно отразиться и от второго опалесцентного пиона, имеющего противоположное направление критического импульса. Таким образом, протон ведет себя подобно фотону в резонансном интерферометре. В результате ядерной реакции, например, поглощения фотона, нуклон может попасть в резонанс с ядерным веществом, обусловленный эффектом критической опалесценции ядра. Этот эффект можно

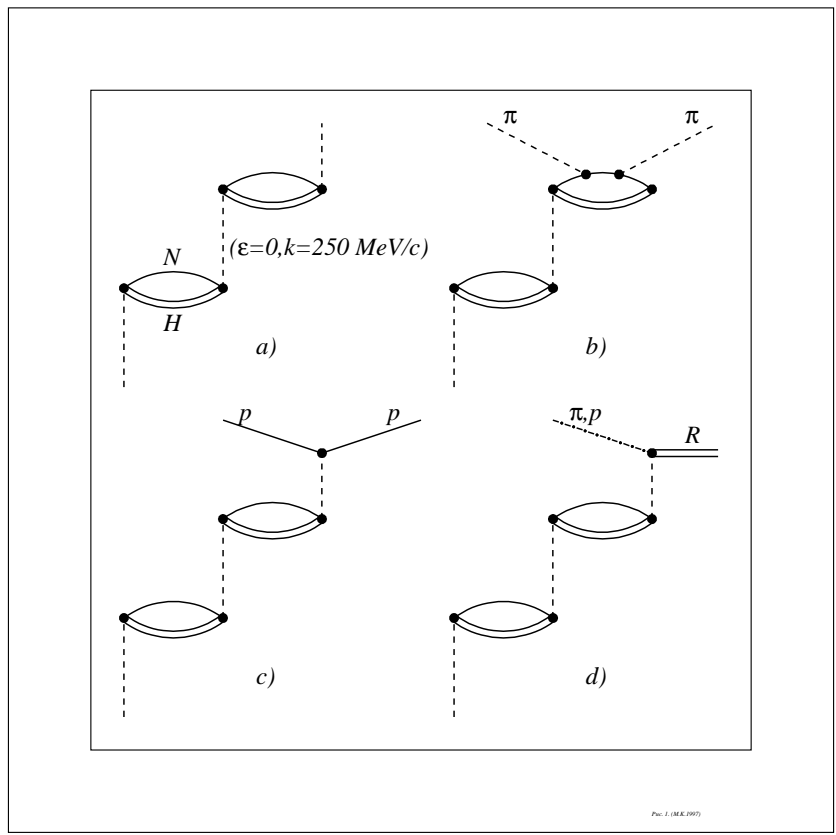

Рис. 15: Диаграммы: $a)$ — опалесцентный пион в ядре, $b)$ — взаимодействие пиона с опалесцентным пионом, $c)$ — взаимодействие протона с опалесцентным пионом, $d)$ — резонансное взаимодействие с опалесцентным пионом.

сопоставить с возбуждением гигантского резонанса в ядре. Уровни гигантского резонанс вдвое выше, чем 8 МэВ, поскольку необходимо затратить дополнительную энергию на преодоление энергии связи нуклона в ядре. Более высокие уровни гигантского резонанса обусловлены несколькими протонами в "опалесцентном интерферометре".

Аналогичный "опалесцентный интерферометр" может захватывать пионы с импульсом 125 МэВ/$c$ (рис. 15$b$). Рассеяние на одном из опалесцентных пионов может объяснить эффект **ядерной полупрозрачности (translucence)**. Сечение взаимодействия пионов с нуклонами в области энергий ниже (3,3)-изобары почти на порядок ниже, чем в области (3,3)-изобары. Если пион с кинетической энергией 200 МэВ имеет длину свободного пробега в ядре порядка нескольких десятых ферми, то пион с кинетической энергией 50 МэВ имеет длину свободного пробега порядка нескольких ферми. В этом смысле ядерную материю можно считать относительно прозрачной для мягких пионов. С другой стороны, в области мягких пионов ($T_\pi$ = 50 МэВ) был обнаружен целый ряд аномальных явлений. Первая аномалия при этой энергии была обнаружена в рассеянии пионов на ядре углерода с выделением $0^+$- и $2^+$-возбуждений ядра в конечном состоянии ядра [93]. Качественно обнаруженный эффект выглядел как взаимодействие пиона с ядром, при котором значительно возрастает вероятность рассеяния пиона на большие углы при относительно низком сечении поглощения, что можно сравнить с рассеянием света в полупрозрачном стекле. Позже полоса полупрозрачности при той же энергии была найдена в энергетической зависимости реакций с перезарядкой пиона $^{15}N(\pi^+, \pi^0)^{15}O$ и $^7Li(\pi^+, \pi^0)^7Be$ [94]. В конце 80-х годов аномалия при энергии пионов 50 МэВ была найдена в виде усиления сечения двойной перезарядки пионов на ядрах [95]. Гипотеза резонансного "опалесцентного интерферометра" позволяет, по крайней мере качественно, объяснить эти эффекты. В частности, двойная перезарядка пионов с импульсом 125 МэВ/$c$ ($T_\pi = 50$ МэВ) может быть усилена, поскольку такой пион может отражаться от обоих "зеркал интерферометра", каждый раз перезаряжаясь.

На резонансные свойства ядерной полупрозрачности впервые обратил внимание У. Гиббс, заметив, что полупрозрачность имеет вид относительно узкой полосы и фаза оптического потенциала, описывающего рассеяние пиона ядром, близка к $90^o$, поэтому формально "ядерную полупрозрачность" можно было бы интерпретировать как "пион-ядерный резонанс" [96]. Однако существует много других подходов к объяснению аномального эффекта усиления двойной перезарядки пионов при энергии 50 МэВ [97]. Одним из альтернативных объяснений эффекта является гипотеза существования локального $\pi NN$-резонанса ($d'$-дибариона) с массой 2062 МэВ [98]. Указание на существование такого резонанса было получено в ИТЭФ в реакции $p(p, \pi^+)pp\pi^-$ [14]. Возбуждение "пион-ядерного резонанса" в результате взаимодействия частиц высокой энергии с ядром может проявляться как максимум в спектре мягких ($T_\pi = 50$ МэВ) пионов. Указание на существование такого максимума было получено при анализе спектров пионов в ядро-ядерных взаимодействиях [99] задолго до обнаружения ядерной полупрозрачности при той же энергии. В ИТЭФ аномалия в энергетическом спектре пионов была обнаружена в пион-ядерных реакциях [100, 101]. В [100] усиление выхода пионов с импульсом 125 МэВ/$c$ ($T_\pi = 50$ МэВ) было обнаружено при взаимодействии пиона с энергией 3 ГэВ с ядрами $C$ и $Xe$ при отборе событий с единственным вторичным пионом, испущенным в заднюю полусферу (рис. 16а).

Указание на существование эффекта критической опалесценции в ядре было получено в трех экспериментах, проведенных в ИТЭФ. В первом эксперименте изучалась реакция ($\pi, \pi\pi$) на ядре. Были отобраны события с малой переданной ядру энергией ($\Delta E < 50$ МэВ), но сравнительно большим переданным импульсом [101]. После вычитания вклада центральных взаимодействий с ядром и реакций на квазисвободном нуклоне в импульсном спектре виртуальных ядерных пионов проявился максимум. Соответствующие события можно было интерпретировать как взаимодействие с опалесцентным пионом. Для средних ядер ($Ti$, $Fe$) импульс опалесцентных пионов был найден равным 249.4 МэВ/$c$ (рис. 16б). Во втором эксперименте при измерении энергетической зависимости выхода мягких мезонов ($T_\pi = 30$–70 МэВ) было обнаружено (рис. 16в), что в области импульсов налетающего пиона а) Спектр положительных пионов, вылетающих назад при взаимодействии пионов с энергией 3 ГэВ с ядрами $C$ и $Xe$, при отборе событий с единственным вторичным пионом [100] и несколькими протонами.

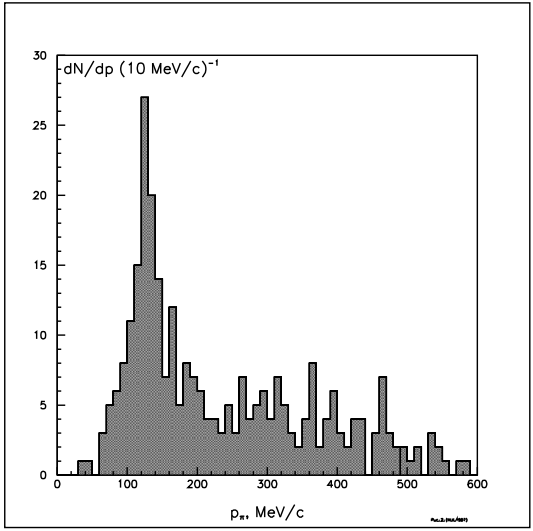

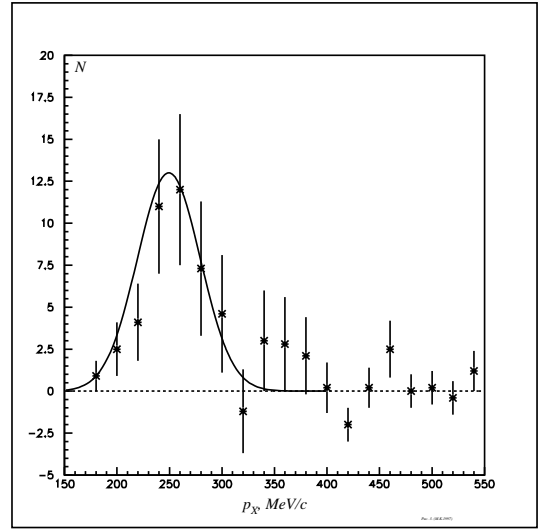

б) Измерение критического импульса опалесцентного мезона в реакции $(\pi, \pi\pi)$ на средних ядрах $(Ti, Fe)$ [101].

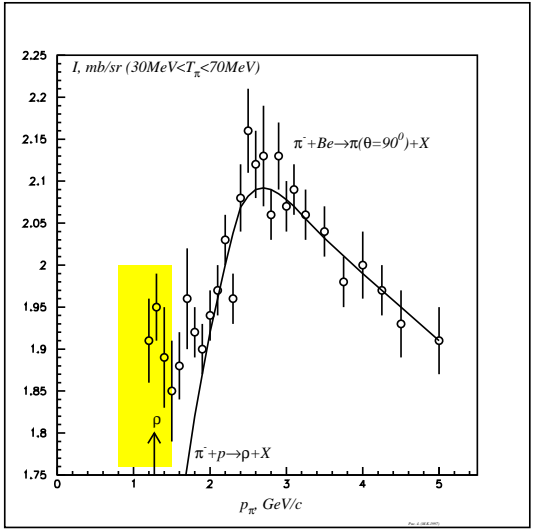

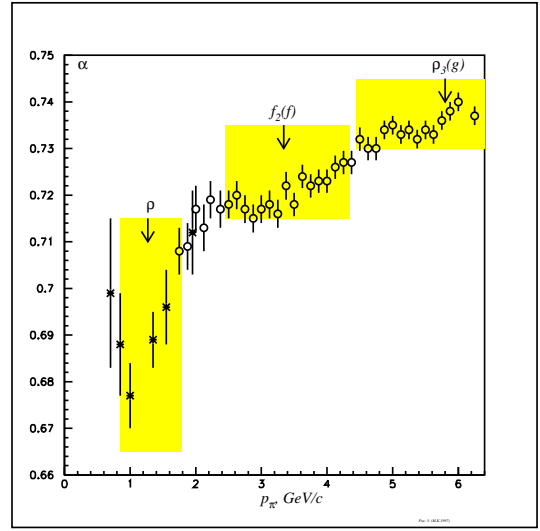

в) Энергетическая зависимость выхода мягких пионов в реакции $Be(\pi, \pi)X$, измеренного под углом $90^o$.

г) Экспериментальный сигнал взаимодействия пиона с опалесцентным пионом ядра, полученный в эксперименте, исследовавшем $A$-зависимость полных неупругих адрон-ядерных сечений [102].

Рис. 16:

вблизи $p_\pi = 1.27$ ГэВ/$c$ выход мягких мезонов значительно превышает ожидаемую величину, рассчитываемую как результат рождения $\rho$-мезонов на квазисвободных нуклонах. Наконец, при переобработке эксперимента, в котором измерялась $A$-зависимость полных неупругих сечений взаимодействия пионов с ядром [102], было получено указание на существование резонансного взаимодействия с опалесцентными пионами не только в области $\rho$, но и в области более высоких $\pi\pi$-резонансов. Измеренная $A$-зависимость полных неупругих сечений пионов с ядром имела степенной характер. Энергетическая зависимость показателя степени $A$ обнаружила нерегулярности, которые можно интерпретировать как взаимодействие налетающего пиона с опалесцентным пионом ядра, увеличивающим полное неупругое сечение взаимодействия пионов с легкими ядрами. При энергиях, соответствующих положению $\pi\pi$-резонансов, наблюдалось снижение показателя степени $A$-зависимости (рис. 16г).

Усиление выхода мягких пионов ($T = 40$–70 МэВ) по отношению к жестким пионам ($T > 100$ МэВ), возникающим в результате распада $\Delta$-изобары, исследовалось в протон-ядерных взаимодействиях в Дубне [103] и на зарубежных ускорителях [104]. В недавней работе [105] было показано, что усиление выхода мягких мезонов имеет сильную зависимость от энергии налетающего протона. При энергии протонов 350 МэВ под углом $90^o$ обнаружено увеличение вдвое отношения выхода мягких мезонов к выходам пионов с энергиями порядка 100 МэВ.

Эффекты, являющиеся указанием на существование ядерной критической опалесценции, требуют дополнительных исследований. Необходимо исследовать энергетическую зависимость выхода мягких пионов с энергией 50 МэВ на пучках пионов и протонов. Диапазон пионных пучков должен перекрывать область от 0.8 до 1.5 ГэВ/$c$, а диапазон протонных пучков должен перекрывать область от 0.9 до 2.0 ГэВ/$c$. Это относительно дешевые инклюзивные измерения, в которых можно использовать сцинтилляционные счетчики типа КОРД [16], установленные на базе 2–2.5 метра. Достаточно исследовать три ядерные мишени $C$, $Fe$, $Pb$.

## 3.5. Исследование "выброса" ядерной материи в ядро-ядерных взаимодействиях при энергии ядра-снаряда до 3–4 ГэВ/$N$

Предлагается изучение "выброса" ядерной материи (испускание многозарядных фрагментов) в направлении, перпендикулярном к плоскости реакции. В опытах на установке Plastic Ball [106] было показано, что плоскость ядерной реакции может быть определена направлением пучка и бо́льшей осью трехмерного эллипсоида, описывающего поток энергии. Это позволяет суммировать все взаимодействия таким образом, чтобы совпали их плоскости реакции. В этом случае на большой статистике взаимодействий было показано, что в азимутальном распределении (относительно вектора, направленного вдоль бо́льшей оси эллипсоида) наблюдается асимметрия вылета частиц. Величина отношения

$$R_N = \frac{N(+90^o) + N(-90^o)}{N(0^o) + N(180^o)}$$

количества частиц, вылетевших перпендикулярно плоскости реакции, к их потоку в плоскости реакции показана для различных ядер на рис. 17 в зависимости от нормированной множественности частиц в средней области псевдобыстрот. Как видно из этих данных, эффект асимметрии сильно зависит от массы снаряда.

Был измерен "выброс" для одно- и двухзарядных фрагментов. Причем наблюдался большой рост асимметрии для $\alpha$-частиц. Исследование проведено только до энергии 1.2 ГэВ/$A$ без выделения фрагментов с зарядом больше 2. Результаты измерений показаны на рис. 18. Видно, что наблюдается сильная зависимость от массы (заряда) фрагмента. На рис. 19 показано азимутальное распределение для средней области псевдобыстрот и различных величин прицельного параметра. MUL1 соответствует периферическому, а MUL5 – центральному соударению. Наибольший эффект наблюдается у тяжелого ядра-снаряда при средней множественности, соответствующей "боковому" соударению ядер (MUL3 и MUL4).

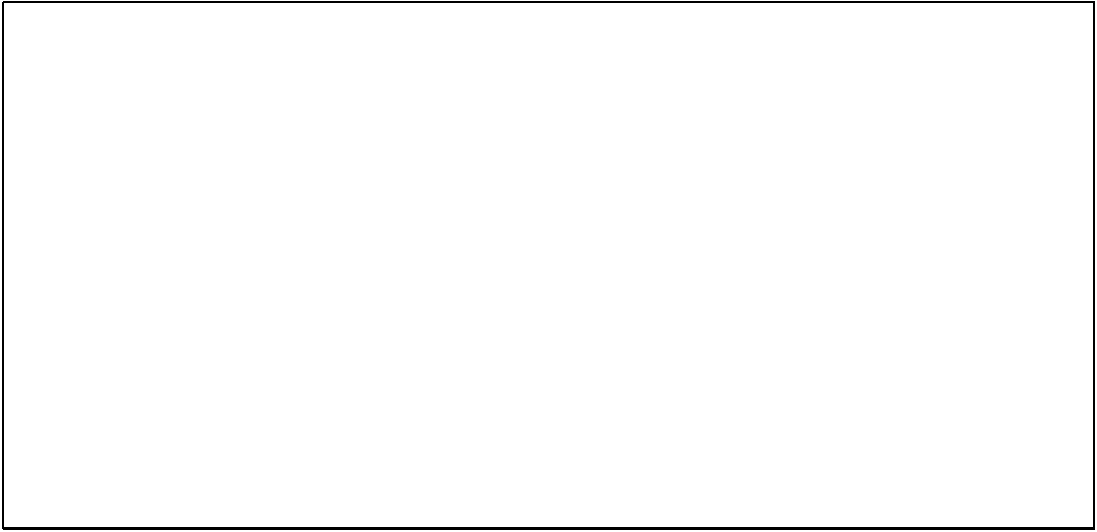

Рис. 17: Зависимость величины $R_N$ от нормированной множественности для различных ядер-снарядов при энергии 400 МэВ/нуклон.

Рис. 18: Зависимость величины $R_N$ от заряда фрагментов для ядра-снаряда $Au$ при энергии 400 МэВ/нуклон.

Предлагается методом ядерных фотоэмульсий изучить "выброс" многозарядных ядерных фрагментов при энергии различных налетающих ядер-снарядов до $3 \div 4$ ГэВ/$N$. Метод обладает высоким пространственным разрешением, $4\pi$-геометрией и сравнительной простотой определения зарядов ядерных фрагментов. В настоящий момент разработана методика определения зарядов с помощью микроскопа, связанного в линию с ЭВМ. На рис. 20

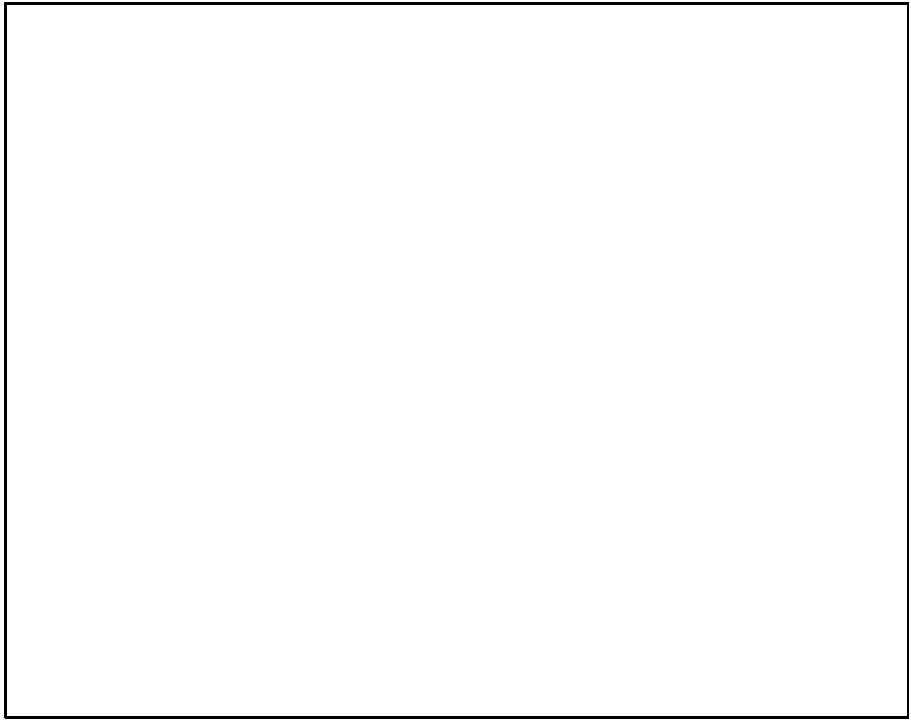

Рис. 19: Азимутальное распределение относительно осей потока энергии для различных ядер-снарядов и множественностей.

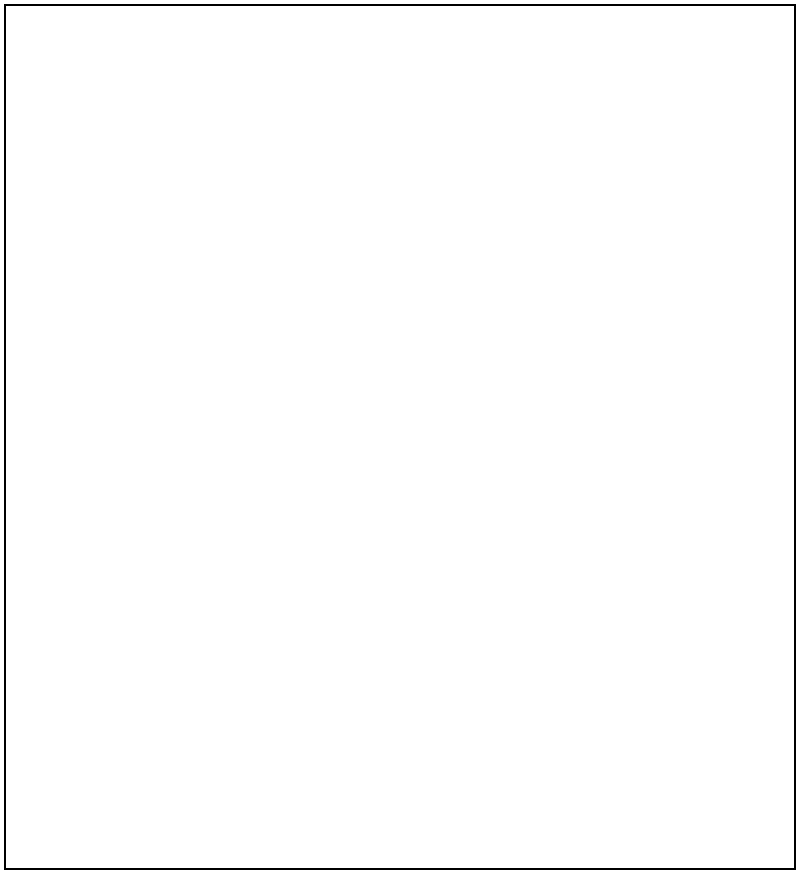

Рис. 20: Зависимость числа просветов с длиной больше $l_i$ от заряда частицы.

приведены калибровочные измерения определения заряда по углу наклона интегрального распределения длин просветов, а на рис. 21 — нормированное значение для просветов больше 0.25 мкм в зависимости от величины заряда. Как видно из двух проведенных калибровок, имеется разработанная методика измерения зарядов легких фрагментов до $Z = 5 \div 6$.

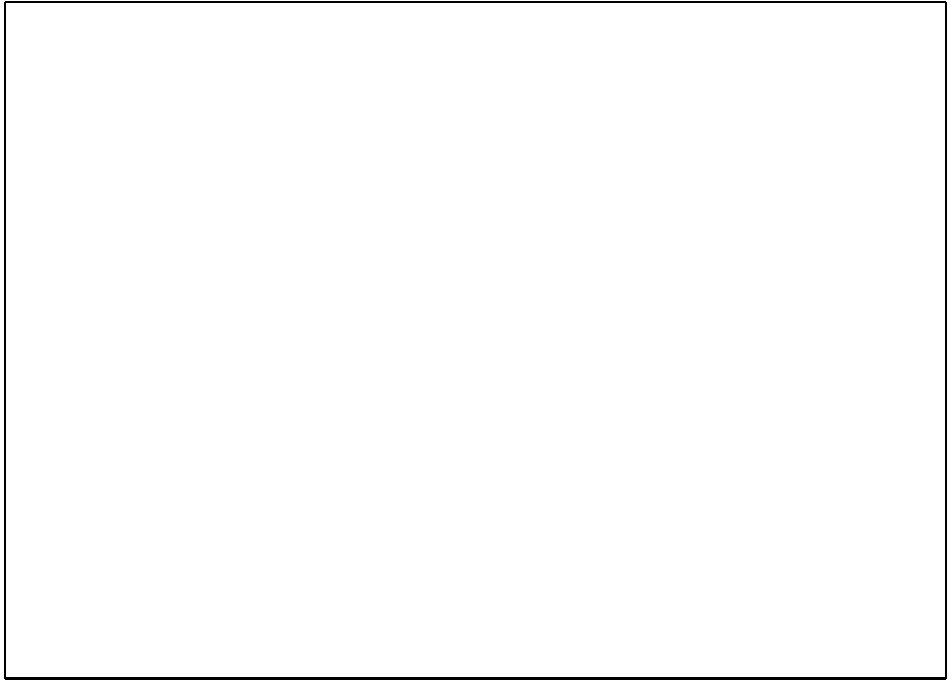

Рис. 21: Зависимость нормированной суммарной длины просветов больше $l = 0.25$ мкм от заряда частиц.

Исследования "выброса" фрагментов в ядро-ядерных взаимодействиях позволит "заглянуть через окно" в горячую область плотной ядерной материи, наблюдая рождение частиц, не замазанное их прохождением через вещество мишени-снаряда. Схематически эта картина показана на рис. 22.

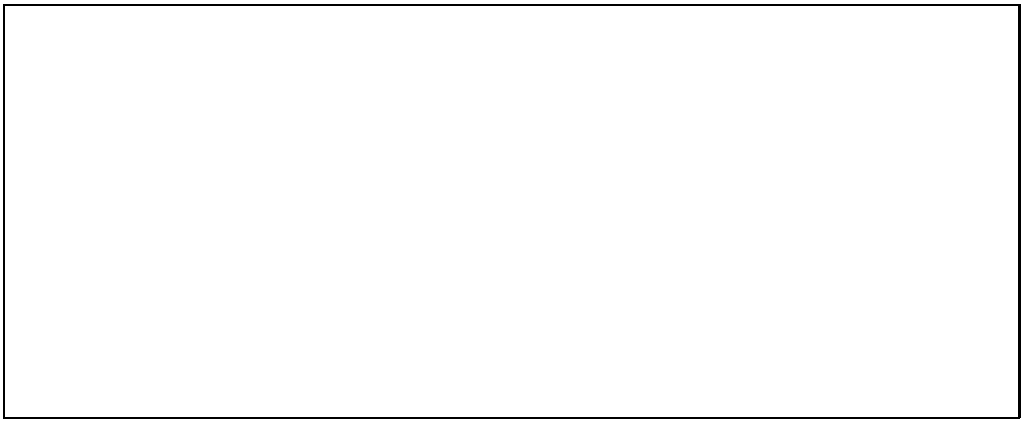

Рис. 22: "Выброс" материи в ядро-ядерных взаимодействиях.

### 3.6. Эффекты многопионных промежуточных состояний в двойной перезарядке пионов на ядрах при энергии выше 1 ГэВ

В исследовании двойной перезарядки пионов на ядрах (ДПП) в ИТЭФ в 1995-1997 годах была обнаружена сильная аномалия в энергетической зависимости сечения этого процесса при $T_0 > 0.6$ ГэВ в области, кинематически запрещенной для реакции $(\pi, 2\pi)$ [107, 108]. При ожидаемом резком падении этого сечения с энергией, измерения дали относительно слабую энергетическую зависимость процесса ДПП с превышением почти на порядок величины наблюдаемого выхода пионов над теоретическим, который получен в предположении об обычном механизме двух последовательных перезарядок на двух нуклонах. Теоретический анализ [109] первых данных по ДПП при энергиях $\gtrsim 1$ ГэВ показал, что процесс ДПП предоставляет уникальную возможность наблюдать эффект *неупругих* глауберовских перерассеяний как основной, уже начиная с $\sim 0.7$ ГэВ. Действительно, в области этих энергий обычный двухступенчатый механизм ДПП с однопионным промежуточным состоянием не доминирует из-за малости сечения однократной перезарядки на нуклоне. Таким образом, целью предлагаемого эксперимента является исследование неупругих механизмов многократного взаимодействия пионов с нуклонами при их прохождении через ядро. При этом намечается получить информацию об энергетическом ходе инклюзивной ДПП вперед на ряде ядерных мишеней при начальном импульсе в интервале $1 \div 3$ ГэВ/$c$.

Известно, что учет глауберовских перерассеяний при прохождении адронов через ядро дает существенные поправки к сечениям соответствующих процессов. Что касается ДПП, то в рамках двухступенчатого механизма двух последовательных однократных перезарядок на двух нуклонах ядра этот процесс тождествен картине *упругого* глауберовского перерассеяния (с $\pi^0$-мезоном в промежуточном состоянии). В обычных адрон-ядерных реакциях вклад упругого перерассеяния может составлять до $\sim 15\%$, а неупругие поправки становятся существенными только при высоких энергиях ($> 5$ ГэВ) и составляют $20 \div 30\%$ от упругих поправок [110]. Для ДПП *неупругие* перерассеяния (с многопионными промежуточными состояниями), как показал развитый А.Б. Кайдаловым на основе грибовского формализма [111] подход к объяснению обнаруженной аномалии, являются *доминирующим* механизмом, вклад которого возрастает с энергией. Теоретические оценки показывают, что при энергиях $1\div3$ ГэВ можно ожидать продолжения слабого хода сечения ДПП в условиях рассматриваемого эксперимента. Таким образом, в рамках планируемого относительно простого эксперимента представляется возможным не только подтвердить факт сильного доминирования неупругих перерассеяний в процессе ДПП, но и впервые измерить ход сечения ДПП в интервале $1 \div 3$ ГэВ. Программа изучения ДПП при ГэВ-ных энергиях на других ускорителях отсутствует; в качестве заявки она сформулирована только для SIS (GSI, Дармштадт) с участием ИТЭФ. Отсутствие подходящей установки позволяет надеяться, что удастся сохранить лидерство в изучении ДПП в ИТЭФ.

### 3.7. Изучение векторных ($\rho$, $\omega$, $\varphi$) и скалярных ($\eta$, $\eta'$) мезонов, образованных при взаимодействии пионов с тяжелыми ядрами

Вопросы о свойствах адронных резонансов в ядерной среде являются в последние годы предметом оживленной дискуссии. Модели эффективного лагранжиана [112, 113, 114] и подходы, основанные на правилах сумм КХД [115, 116] предсказывают уменьшение масс $\rho$, $\omega$, $\varphi$ с увеличением ядерной плотности. Кроме того, с уменьшением масс фазовый объем для распадов резонансов также уменьшается, что приводит к увеличению ширины резонансов. С другой стороны, благодаря столкновениям, зависящим от плотности ядерной среды и сечения взаимодействия резонанса с нуклоном [117, 118], ширина резонанса также увеличивается.

Свойства векторных мезонов в ядерной среде изучались экспериментально путем измерения дилептонов на ускорителе SPS (CERN) как для протон-ядерного, так и для ядро-ядерного взаимодействий [83, 119, 120, 121].

В работе Li и др. [122] высказано предположение, что избыток в спектре дилептонных масс в районе $0.3 \leq M \leq 0.7$ ГэВ во взаимодействиях $S + Au$ по сравнению с $p + Au$ может быть объяснен сдвигом массы $\rho$-мезона.

Приведенный в работах [85, 123, 124] анализ также указывает на такую возможность, однако не исключает более простую интерпретацию за счет "self-energy" эффекта в ядерной материи.

Учитывая неопределенность экспериментальной ситуации и важность проблемы, связанной с проверкой предположения о частичном восстановлении киральной симметрии в ядерной среде, являющейся фундаментальной симметрией КХД, необходимо провести дополнительные эксперименты и получить информацию о свойствах векторных (и скалярных) мезонов в более ясных динамических условиях, например, в пион-ядерных столкновениях.

Особенно заметные изменения характеристик адронов предсказываются для адронов, рожденных в сверхплотной перегретой ядерной материи, образующейся, например, при столкновении ускоренных до нескольких ГэВ на нуклон тяжелых ядер с тяжелыми ядрами.

Однако и при рождении адронов фотонами, электронами или пионами на тяжелых ядрах, т.е. при нормальной плотности ядерной материи ($0.17$ Фм$^{-3}$), следует ожидать хотя и меньших ($\Delta m/m \approx 10 \div 20\%$), но вполне измеримых эффектов. Более того, в некотором отношении такие процессы предпочтительнее, поскольку позволяют проследить изменение параметров рожденных адронов на протяжении всего времени их жизни без временных ограничений, обусловленных малым временем жизни сверхплотной перегретой ядерной материи. Кроме того, реакции, вызванные пионами, позволяют вести исследования при меньшем комбинаторном фоне по сравнению с протонами и тяжелыми ядрами. В этих реакциях возможна более прямая и менее модельно зависимая интерпретация данных.

Адроны, могущие служить в качестве "пробников" для наблюдения ожи-даемых эффектов, должны удовлетворять двум критериям:

1) иметь достаточно малое время жизни, чтобы распасться внутри ядра;

2) иметь каналы распада, не искаженные сильным взаимодействием распадных частиц, т.е. лептонные и электромагнитные распады.

Этим условиям удовлетворяют векторные и скалярные мезоны, приведенные в таблице 2.

Таблица 2: Относительные вероятности распадов мезонов

| | Каналы распада | | | | | | |
|---|---|---|---|---|---|---|---|
| | $e^+e^-$ | $\mu^+\mu^-$ | $\mu^+\mu^-\gamma$ | $\pi^0\gamma$ | $\gamma\gamma$ | $\eta\gamma$ | $3\gamma$ |
| $\eta$ | $< 7.7 \cdot 10^{-5}$ | $5.8 \cdot 10^{-6}$ | — | — | $39 \cdot 10^{-2}$ | — | — |
| $\eta'$ | $< 2.1 \cdot 10^{-7}$ | — | $1 \cdot 10^{-4}$ | — | $2.1 \cdot 10^{-2}$ | — | — |
| $\rho^0$ | $4.5 \cdot 10^{-5}$ | $4.6 \cdot 10^{-5}$ | — | $6.8 \cdot 10^{-4}$ | — | $2.4 \cdot 10^{-4}$ | $2 \cdot 10^{-4}$ |
| $\omega^0$ | $7.1 \cdot 10^{-5}$ | $< 1.8 \cdot 10^{-4}$ | — | $8.5 \cdot 10^{-2}$ | — | $6.5 \cdot 10^{-4}$ | $< 1.9 \cdot 10^{-4}$ |
| $\varphi^0$ | $3.0 \cdot 10^{-4}$ | $2.5 \cdot 10^{-4}$ | $2.3 \cdot 10^{-5}$ | $1.3 \cdot 10^{-3}$ | — | $1.26 \cdot 10^{-2}$ | — |

Представляется, что измерения следует вести либо по радиационным (там, где мода распада велика), либо по мюонным распадам. Выделение электронных распадов потребует изготовления сложного и дорогого счетчика RICH и, кроме того, будет затруднено из-за большого фона быстрых пионов. При регистрации мюонных распадов можно ожидать существенно более благоприятных фоновых условий. При использовании пионных пучков с импульсами до 2 ГэВ/$c$ образование векторных мезонов будет происходить, в основном, в следующих элементарных процессах:

$$\pi^- p \;\to\; \omega n \,,$$
$$\pi^- p \;\to\; \omega \pi N \,,$$
$$\pi^- p \;\to\; \rho n \,,$$
$$\pi^- p \;\to\; \rho \pi N \,,$$
$$\pi^- p \;\to\; \varphi n \,.$$

Сечение образования векторных мезонов при $p_{\pi^-} < 2$ ГэВ/$c$ составляет величину $\sim 2$ мб для $\omega$- и $\rho$-мезонов и 20 мкб для $\varphi$-мезона [125].

В работе [125] рассчитан сдвиг масс и уширение $\omega$-мезона в ядерной среде, в частности, в реакции $\pi^- + {}^{208}Pb$ при $1.1 \leq P_{\pi^-} \leq 1.7$ ГэВ/$c$. В ней обращается внимание на необходимость отбора малых продольных и поперечных импульсов $\omega$-мезона ($P_L$, $P_T \leq 0,25$ ГэВ/$c$), с тем, чтобы лучше выделить влияние ядерной среды. Результаты расчета приведены на рис. 23.

Следует обратить внимание на следующий эффект. Если эффективная масса $\varphi$-мезона в ядерной материи, как ожидается в ряде теоретических работ, уменьшится на величину 30 МэВ, энергетически станет невозможным его распад на $K^+K^-$-пару. Поэтому важно наблюдать уменьшение (если таковое окажется) выходов $K^+K^-$-пар во взаимодействиях $\pi^-$+ядро по сравнению с реакцией $\pi^- p \to n\varphi$, в которой $\varphi$-мезон практически покоится.

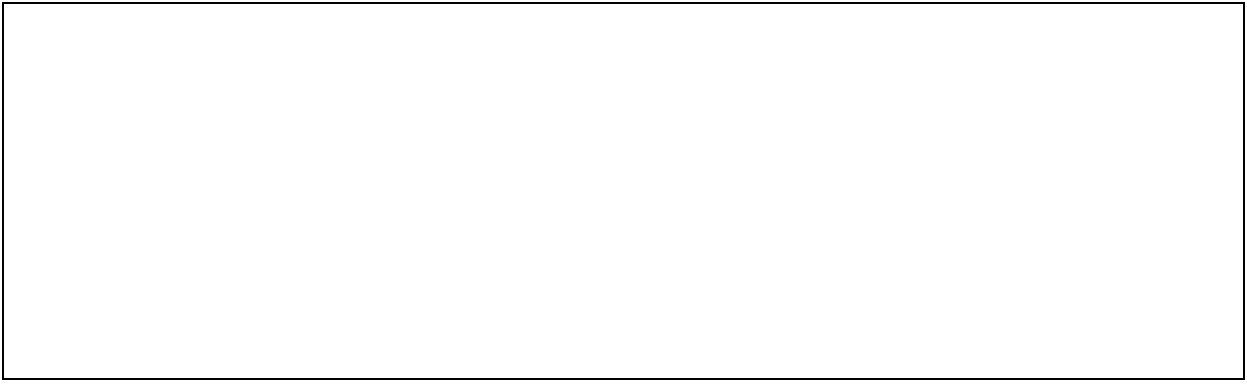

Рис. 23: Инклюзивное дифференциальное сечение образования лептонных пар от прямого распада $\omega \to e^+ e^-$ при различных начальных импульсах с отбором $P_L$, $P_T \leq 0,25$ ГэВ/$c$. Величина $R$ — отношение провзаимодействовавших и непровзаимодействовавших $\omega$-мезонов в ядерной среде.

### 3.8. Адрон-ядерные взаимодействия с рождением систем $pp\pi^-$, $pp$, $pn$ в области, далекой от квазисвободной кинематики

Основной целью исследований является продвижение в район *малых расстояний* между нуклонами в ядре с целью изучения той области, где должен происходить переход от описания процессов на языке адронной физики к описанию на языке КХД. Предполагается исследование как простых механизмов взаимодействия, так и проявлений *экзотических возможностей*.

Предлагается изучение реакции

$$(p, \pi) + A \to (p, \pi)' + A' + pp\pi^- \ (pp, pn, d)$$

на дейтериевой мишени и мишенях из других *легких ядер*.

К кинематическим особенностям эксперимента относятся:

- *большие импульсы* (по сравнению с характерным значением фермиевского импульса) вторичных нуклонов, включая и нуклоны-спектаторы (выше 250 – 300 МэВ/$c$);
- большие переданные импульсы от начального к конечному адрону;
- *широкая область масс* (начиная от значений, близких к порогу) системы вторичных адронов $pp\pi^-$, $NN$.

Основные физические предложения (в порядке возрастания экзотических черт).

1. *Изменение видимых параметров* $\Delta$–изобары и других барионных резонансов $N^*$ *в ядерной среде.* Здесь имеются предсказания, основанные как на модели перерассеяния продуктов распада широких резонансов на остаточной ядерной системе [126], так и, что особенно интересно для более тяжелых ядер, на влиянии ядерной среды, рассматриваемой как конденсированная система [127]. Известны экспериментальные данные, указывающие на подобные эффекты [128].

2. Исследование характерных черт *механизмов взаимодействия, более сложных, чем квазисвободное рождение* пионов. В частности, треугольный механизм с промежуточным образованием $\Delta$–изобары или другого барионного резонанса $N^*$ предсказывает движение пика в распределении по инвариантной массе системы $pp\pi^-$ при изменении импульса, переданного от начальной к конечной быстрой частице. (Эта картина, связанная с наличием так называемых движущихся комплексных особенностей, пока экспериментально не наблюдалась.) Исследования подобного рода интересны не только на дейтроне, но и на других легких ядрах.

3. Детальное исследование системы $pp\pi^-$ для выяснения вопроса о природе *особенности, наблюдаемой в спектре ее инвариантных масс*. Имеются данные группы ИТЭФ [129] и данные группы Тюбинген–Уппсала [130], указывающие на существование узкого дибарионного резонанса $d'$ с массой 2.06 ГэВ и с квантовыми числами, запрещенными для системы $NN$ [131]. На то же указывают и узкие пики в функции возбуждения при 50 МэВ в реакциях двойной перезарядки пионов на легких ядрах [130]. Крайне актуальны исследования в более широкой области энергий на разных ядерных мишенях и с хорошей статистической обеспеченностью.

4. *Поиск и возможное исследование "экзотических" состояний дейтрона*, в первую очередь возможных примесей компонент $\Delta\Delta$, $NN^*$, $N^*N^*$ путем изучения реакций выбивания $\Delta$ и $N^*$. Основной трудностью в исследовании таких компонент, вес которых ожидается в доли процента, является вопрос о возможной имитации за счет процессов образования $\Delta$ и $N^*$ в самом процессе взаимодействия. То новое, что позволит нам продвинуться по сравнению с предыдущими работами - привлечение для четкого выделения процессов истинного выбивания критерия Треймана–Янга и исследования характерного углового распределения продуктов распада резонансов, предсказываемого теорией [132]. Интересно было бы исследовать примеси состояний с $\Delta$ и $N^*$ также и в других легких ядрах.

5. Поиск и возможное исследование "более экзотических" состояний $X$, в которых могут находиться два нуклона в ядре (образ – шестикварковый мешок). Предполагается исследовать рождение системы вторичных адронов ($pp$, $pn$, $d$, $pp\pi^-$) при малых инвариантных массах (вблизи порога) и одновременно в области больших переданных импульсов от налетающей частицы к рожденной системе адронов. Впервые предполагается *попытка прямой идентификации состояний $X$ путем выделения "квазисвободного" механизма выбивания $X$ из дейтрона и других легких ядер с привлечением критерия Треймана–Янга.*

Эксперимент может быть реализован с использованием системы счетчиков КОРД [16].

## 3.9. Исследование структуры легких радиоактивных и стабильных ядер при промежуточных энергиях

В последние годы возрос интерес к исследованию характеристик легких ядер с помощью исследований взаимодействий этих ядер при промежуточных энергиях с протонами и ядрами.Такого типа работы при изучении стабильных ядер типа $^6Li$ обычно проводились, как правило, в прямой кинематике на протонных пучках при использовании ядра в качестве мишени. Например, на хорошей статистике было изучено дифракционное рассеяние протонов с энергией 0.6 и 1 ГэВ на ядре $^6Li$ при относительно малых переданных импульсах, где теория Глаубера-Ситенко достаточно хорошо описывает экспериментальные данные; удалось также достигнуть значений переданных импульсов, при которых возникли существенные разногласия [133]. Полученные в этих работах результаты выявили вопросы, касающиеся как точности и пределов применимости глауберовского метода, так и адекватности используемых волновых функций ядра мишени. В связи с этим возникли предложения по измерению дифракционного рассеяния на симметричном радиоактивном ядре $^6He$ и других экзотических легких ядрах, т.е. ядрах, удаленных от линии $\beta$-стабильности. Естественно, что из-за малых времен жизни этих ядер, такие измерения могут быть выполнены только в обратной кинематике на пучках ядер, которые можно получить методом срыва ускоренных стабильных средних ядер.

С конца 80-х годов в зарубежных институтах для измерений сечений взаимодействий, фрагментации, кулоновской диссоциации, измерения радиусов и структуры нейтрон-избыточных ядер начали создавать пучки радиоактивных ядер и необходимые экспериментальные установки с энергиями ядер до 100 $A$·МэВ (GANIL и RIKEN) и до 1 $A$·ГэВ (Saturn).

В результате проведенных исследований удалось обнаружить ряд необычных свойств легких экзотических ядер (обзоры [134, 135, 136]). Следует особенно отметить, что анализ сечений взаимодействия ядра $^{11}Li$ со стабильными ядрами-мишенями, измерение сечения фрагментации $^{11}Li \rightarrow {}^9Li + 2n$ и измерение импульсного распределения ядер $^9Li$ в этой реакции позволило сделать предположение о существовании в этом ядре нейтронного гало. Есть указания на существование нейтронной "шубы" в ядрах $^6He$, $^8He$, $^{11}Be$ и протонного гало в протон-избыточном ядре $^8B$ [137, 138]. При этом следует различать понятия "шуба" и "гало". Под нейтронной "шубой" понимается состояние, включающее довольно много нейтронов с относительно высокой плотностью, в то время как нейтронное гало содержит один или два нейтрона, крайне слабо связанные и с сильно вытянутым распределением. "Шуба" может проявляться на любой орбите и образуется в радиальной области, когда потенциальная энергия становится приблизительно равной энергиям нескольких наиболее высоких орбит одиночных частиц. Гало рассматривается как чисто туннельный эффект, возникающий, по всей вероятности, только для $s$- и $p$-орбит вследствие центробежного барьера. Обычная плотность нейтронов в нейтронном гало менее 1/100 плотности нейтронов в центральной области. Так как нейтронная шуба более велика по размерам, она может быть отделена от нейтронного гало, хотя они могут иметь и перекрывающиеся области.

В связи с возрастающим интересом к свойствам легких и экзотических ядер предлагается использовать возможность получения экзотических ядер на проектируемом ионном ускорителе ИТЭФ для продолжения исследований свойств таких ядер, как например, $^8He$, $^{10}He$, $^8Li$, $^7Be$, $^9Li$, $^{11}Li$, $^{11}Be$.

Запуск ионного ускорителя ИТЭФ дает возможность создать отдельный выведенный пучок радиоактивных экзотических легких ядер с энергиями, значительно более высокими (до 2000 $A\cdot$МэВ), чем используемые в настоящее время в зарубежных лабораториях, и, таким образом получать, новую информацию о свойствах легких ядер при упругих и неупругих взаимодействиях. На этом пучке можно будет разместить универсальный спектрометр для проведения цикла работ по исследованию угловых распределений упругого и квазиупругого рассеяния и перезарядки легчайших ядер на стабильных и радиоактивных ядрах. Важными являются исследования узких корреляций частиц, вылетающих при взаимодействии ядер. Хорошо также известно, что в легких ядрах четко проявляются кластерные степени свободы и эффекты близости к трехчастичным порогам (например, ядро $^6He$ может рассматриваться как трехтельная система $\alpha + n + n$), что позволяет в хорошем приближении рассматривать легкие ядра как динамические системы, состоящие из нескольких кластеров (альфа-частиц, ядер гелия-3, ядер трития и нуклонов) и рассчитывать структуру ядер путем использования соответствующих детально разработанных методов решения динамических уравнений (многонуклонных уравнений Шредингера, Фаддеева и др.). На установке можно будет проводить следующие исследования.

- Измерения сечений упругого и квазиупругого рассеяния экзотических и радиоактивных ядер на протонах, нейтронах (используя дейтерированный полиэтилен) и легких ядрах при энергиях выше 500 $A\cdot$МэВ, в первую очередь для рассеяния $^6He$ и других экзотических ядер на протоне и нейтроне при больших переданных импульсах:

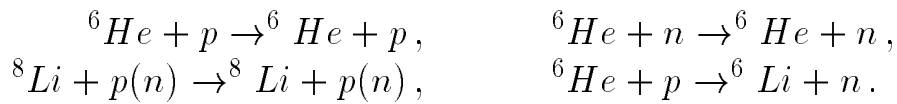

$$^6He + p \to{}^6 He + p\,, \qquad {}^6He + n \to{}^6 He + n\,,$$
$$^8Li + p(n) \to{}^8 Li + p(n)\,, \qquad {}^6He + p \to{}^6 Li + n\,.$$

- Исследования структуры ядер, например, $^6He$ и $^8He$: $^6He + p \to{}^4 He + p + 2n$.
- Исследование узких корреляций вылетающих частиц при развале легких ядер.

● Исследование кластерной структуры легких ядер:

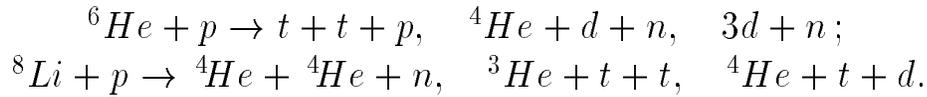

$$^6He + p \rightarrow t + t + p, \quad ^4He + d + n, \quad 3d + n;$$
$$^8Li + p \rightarrow {}^4He + {}^4He + n, \quad ^3He + t + t, \quad ^4He + t + d.$$

Одним из важных открытий в ядерной физике в последние годы было обнаружение расширенного нейтронного распределения в экзотических нейтронно-обогащенных ядрах типа $^{11}Li$, $^{11}Be$, $^8He$, $^6He$. Эти исследования были выполнены, главным образом, с помощью измерений сечений взаимодействий и изучения фрагментации ядер при относительно низких энергиях [139, 140, 141]. В Японии в 1996 г. в RIKEN в режиме обратной кинематики изучено упругое рассеяние радиоактивных ядер $^6He$ и $^3H$ протонами при энергии ядер 71 и 73.5 МэВ, соответственно, и проведены поиски нейтронного гало в ядрах $^6He$ , $^8He$, $^{11}Li$ [139]. Вторичные пучки были получены с помощью сепаратора при фрагментации $^{18}O$ с энергией 100 $A$·МэВ на тонкой $Be$ мишени. При этом, измерения проводились в относительно небольшом диапазоне углов рассеяния протонов (15 ÷ 65 градусов в системе ц.м.).

Существуют теоретические предположения, что такие новые явления как "гало" и "шубы" в экзотических ядрах наиболее ярко должны себя проявлять при больших углах рассеяния и бóльших энергиях (следует отметить, что при бóльших энергиях и теоретический анализ, выполненный в эйкональном приближении, которым пользовались в работе [139], делается более осмысленным). Поэтому важно исследовать экзотические легкие ядра, удаленные от линии $\beta$-стабильности, в как можно большем угловом и энергетическом диапазоне. Насколько известно, такие измерения пока нигде не планируются.

Установка должна содержать следующие элементы:

● выведенный (желательно в корпус 120) пучок ядер среднего атомного номера с временной растяжкой;

● тонкую $Be$ мишень (при невозможности или трудностях с выведением пучка из ускорителя для фрагментации пучка ускорителя можно, но нежелательно, использовать внутреннюю мишень ускорителя);

● магнитную оптику с сепаратором для выделения нужных изотопов ядер;

● отклоняющий магнит типа СП-3 с увеличенным зазором для определения импульсов и углов вылета фрагментов развала исследуемых ядер, образующихся при взаимодействии налетающих ядер с ядрами мишени;

● годоскопы, проволочные пропорциональные камеры, нейтронные детекторы, сцинтилляционные детекторы для регистрации первичных и вторичных частиц, дрейфовые камеры, стрипповые и силиконовые детекторы.

Установку можно монтировать по частям в соответствии с требованиями очередного эксперимента.

## 4. Экспериментальная установка

Серьезным шагом на пути реализации предложенной выше физической программы стало бы создание универсального магнитного спектрометра общего пользования. Предполагается, что набор аппаратуры в таком спектрометре был бы достаточно гибким и позволил решать самый широкий спектр физических задач. Наиболее удачно такая универсальная установка может быть построена на базе модернизированного трехметрового магнита МС-1, имеющегося в лаборатории В.В. Куликова. Модернизация предусматривает увеличение межполюсного зазора магнита в два раза с 50 до 100 см с сохранением основной части магнита и обмоток в прежнем виде. На рис. 24 показан разрез магнита в плоскости, перпендикулярной пучку, после модернизации. Двойная штриховка показывает дополнительные вставки, которые необхо-

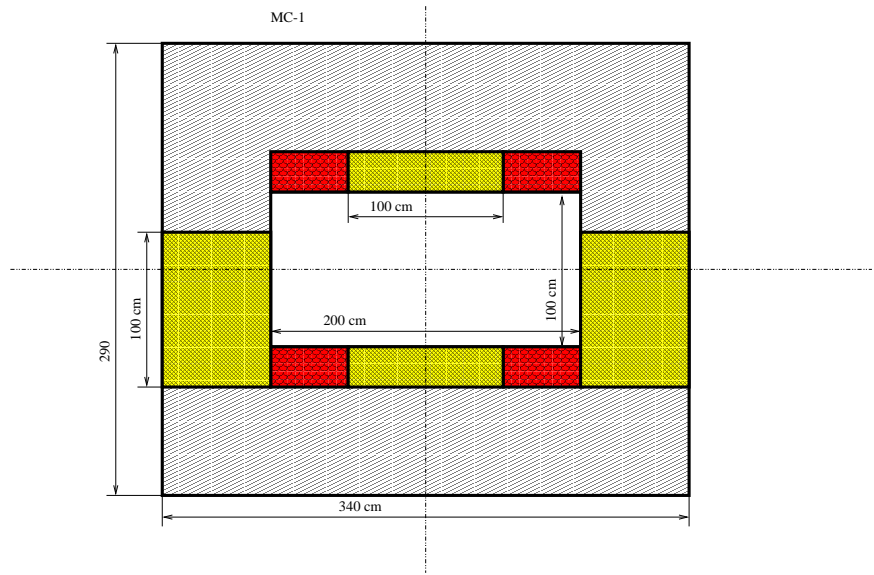

Рис. 24: Вертикальный разрез магнита МС-1 после модернизации димо установить. При сохранении прежнего тока (и потребляемой мощности) поле в центре магнита составит 0.85 Тл[1]. Интеграл поля вдоль пучка составит 2.4 Тл·м, что представляется достаточным для решения большинства поставленных задач. Магнит расположен на универсальном пучке ускорителя ИТЭФ с углом вывода частиц 3.5°. При работе ускорителя в номинальном режиме возможно получение пучков пионов и протонов высокого качества с интенсивностью до $2 \cdot 10^6$ за сброс. Проработана возможность модернизации магнито-оптического тракта для повышения качества фокусировки пучка, что необходимо при работе с поляризованными мишенями. Имеется схема организации медленного вывода ионов в тот же канал.

В качестве трекового детектора представляется разумным использование проволочных дрейфовых камер с длиной дрейфа 1 см. Предполагается иметь 3–4 пучковых камеры размером $20 \times 20$ см$^2$ и 7–10 больших камер с размером $90 \times 170$ см$^2$. Каждая камера должна иметь пять слоёв: 2–3 с вертикальными проволочками и 2–3 стереослоя с проволочками под углом $10^o \div 15^o$ к вертикали. Камеры должны иметь пространственное разрешение не хуже 0.2 мм вдоль горизонтальной оси и 1 мм вдоль вертикальной. Это даст точность измерения импульса 0.4% для частицы с импульсом 2 ГэВ/$c$ и длиной трека 2 м и 0.8% для частицы с импульсом 200 МэВ/$c$ и длиной трека 0.5 м. Наличие отдельных камер обеспечит необходимую модульность аппаратуры и возможность её различного конфигурирования в различных экспериментах.

---

[1]Расчеты трехмерной модели поля, которые ведутся сейчас под руководством В.В. Рыльцова, демонстрируют неоднородность поля между полюсами около 25%, что представляется вполне удовлетворительным.

Для организации триггера и идентификации частиц по времени пролёта можно использовать разработанную в ИТЭФ систему КОРД [16]. Эта система имеет временное разрешение 200 пс, что обеспечивает надёжное разделение пионов и протонов на пролётной базе 2 м с импульсами до 2.5–3 ГэВ.

В работе установки можно будет использовать широкий набор мишеней, в том числе жидководородную и жидкодейтериевую мишени, поляризованную протонную и поляризованную дейтериевую мишени с замороженной поляризацией. Создание этих мишеней облегчается наличием рядом с магнитом МС-1 мощных криогенных инфраструктур лабораторий 301 и 305, а также огромным опытом указанных лабораторий в создании и эксплуатации криогенных и поляризованных мишеней. Объём жидко-газовых мишеней составит $(1 \div 1.5) \cdot 10^3$ см$^3$, поляризованных — 60–80 см$^3$.

На рис. 25 показан вариант расположения аппаратуры с использованием поляризованной протонной мишени с замороженной поляризацией для изме-

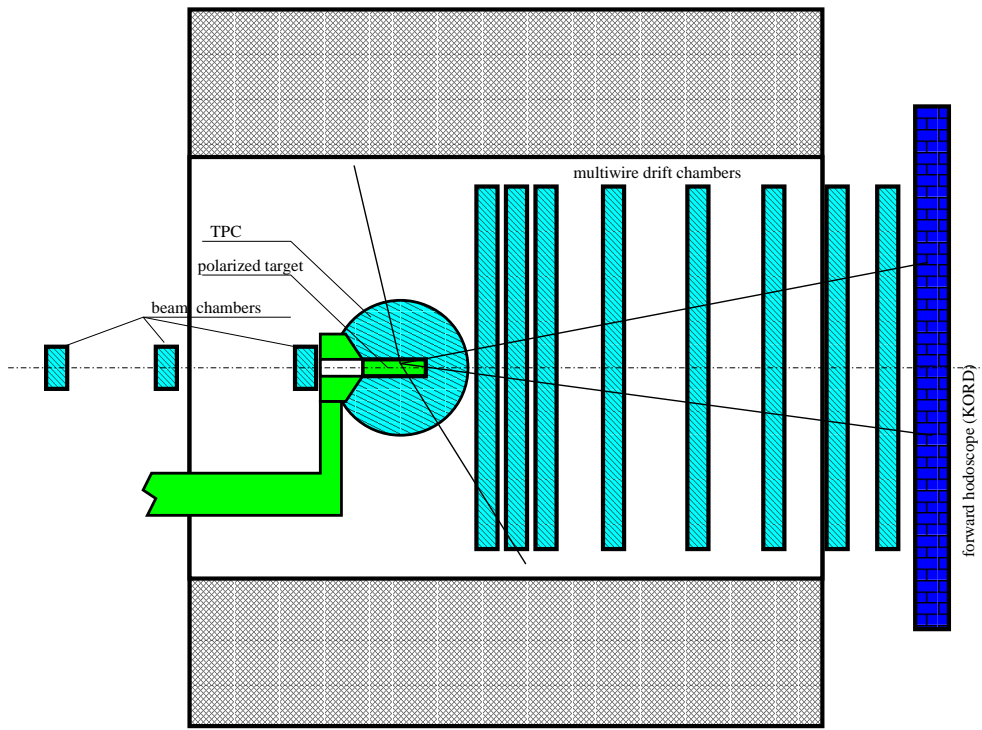

Рис. 25: Схема эксперимента по измерению спин-зависящих переменных в реакциях с рождением странности рения спин-зависящих переменных в реакциях с рождением странности.

Электроника установки должна обеспечивать оцифровку, анализ и накопление не менее 2 тысяч триггеров за сброс и предусматривать возможность организации многоуровневого триггера. На рис. 26 показана схема организации электроники "в линию", составленная в основном из стандартных, а, следовательно, максимально недорогих элементов, таких как компьютеры IBM PC, и связь Ethernet. В настоящее время в лаб. 305 ИТЭФ начата разработка блоков ВЦП с разрешением 0.5–0.7 нс для дрейфовых камер в стандарте FAST CAMAC на базе кристаллов TDC-32, разработанных в CERN.

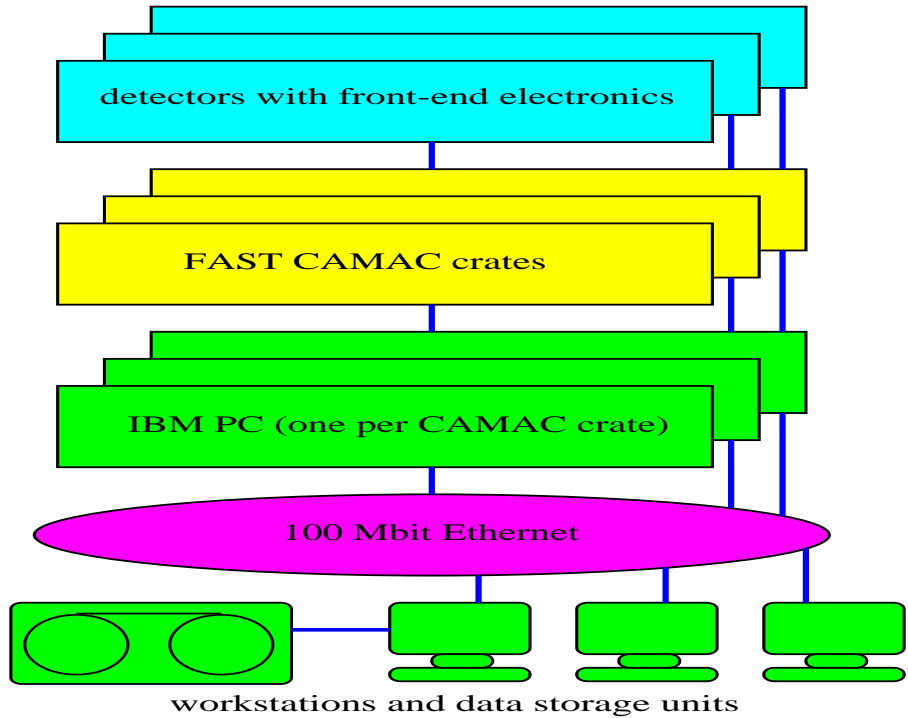

Рис. 26: Схема электроники сбора и накопления информации

Создание установки должно происходить в несколько этапов. На первом этапе планируется модернизировать магнит МС-1, создать дрейфовые камеры, задний годоскоп на базе счетчиков КОРД и жидководородную мишень. Этап рассчитан на три года, в конце которых можно будет уже выйти на физические измерения. На следующем этапе планируется создание поляризованной протонной мишени, разработка центрального детектора на основе времяпроекционной камеры, в которой дрейф будет осуществляться в вертикальном направлении (вдоль силовых линий магнитного поля) и электромагнитного калориметра площадью 5–6 м$^2$ из блоков, имеющихся в ИТЭФ (лаб. 212). Этот этап рассчитан ориентировочно на следующие три года и должен проходить параллельно с получением физических данных на установке, созданной на первом этапе. В это же время или в дальнейшем возможна реализация медленного вывода ионов, создание адронного калориметра, необходимого для решения многих задач по исследованию ядро-ядерных столкновений, и мюонного идентификатора. Одновременно с разработкой аппара-туры следует вести создание программного обеспечения как для работы "в линию", так и для последующей обработки записанной информации, а также провести весьма полное Монте-Карло-моделирование каждой из поставленных задач.

## 5. Заключение

Предварительная программа исследований, представленная выше в виде предложений конкретных экспериментов, свидетельствует о том, что ускоритель ИТЭФ может быть эффективно использован для решения ключевых проблем физики промежуточных энергий. Известно, что основная цель физики промежуточных энергий — понять сильное взаимодействие на расстояниях $(0.1 \div 1)$ Фм в непертурбативном режиме КХД, включая проблему конфайнмента. Эта энергетическая область является переходной от интервала низких энергий, где для описания взаимодействий успешно используются мезонно-обменные модели, к интервалу высоких энергий и больших переданных импульсов, в котором взаимодействие определяется жестким рассеянием индивидуальных партонов (кварков). Базовые проблемы физики промежуточных энергий включают в себя вопросы о природе непертурбативных адрон-адронных взаимодействий, о справедливости мезонно-обменных моделей в странном секторе, где сильно нарушена киральная симметрия, о существовании аномальных мезонных и барионных резонансов, а также дибарионов, о вкладе странных кварков в структуру нуклона, о сечениях взаимодействия резонансов с нуклонами и гиперонов и $\Delta$-изобар друг с другом, о модификации адрон-адронных взаимодействий в ядерной среде, о роли ненуклонных степеней свободы в ядерной материи и другие. Энергетическая область $(1 \div 10)$ ГэВ, перекрываемая синхротроном ИТЭФ, оптимальна для экспериментального изучения большинства перечисленных проблем. Представленная выше программа исследований затрагивает многие из этих ключевых вопросов, хотя и не является исчерпывающей.

Поиски экзотических состояний, в особенности $E_{55}$-бариона и $d'$-дибариона, глюболов и гибридов, в том числе тщательный анализ спектра возбужденных состояний нуклона, возможно, вскроют причины ограничения кваркового состава наблюдаемых адронов простыми комбинациями наивной кварковой модели. По сравнению с высокими энергиями, в интервале ускорителя ИТЭФ следует ожидать существенно больших сечений генерации экзотических объектов.

Предлагаемые детальные исследования низколежащих скалярных мезонов, включающие измерение спиновых параметров в реакциях рождения, прояснят вопрос о составе основного нонета скалярных мезонов, до настоящего времени не установленного окончательно. Особенно интересна до сих пор открытая проблема существования и характеристик скалярно-изоскалярного сигма-мезона, являющегося необходимым объектом современной киральной теории возмущений (LSM-линейные сигма-модели). В них этот мезон определяет массы мезонных резонансов и нуклона. Столь же важны исследования генерации и распадов $f_0$- и $a_0$-мезонов, которые можно рассматривать как необычные адронные состояния (например, "вакуумные скаляры" [18]).

Проверка правила запрета OZI важна для уточнения теоретических представлений о механизме рождения частиц и, возможно, оценки содержания странных кварков в нуклоне.

Несомненный интерес представляет предложение об исследовании радиационных распадов векторных и псевдоскалярных мезонов, вероятности которых зависят от магнитных моментов кварков и кваркового состава мезонов и могут быть сопоставлены с теоретическими предсказаниями (так же как и поиски распада $\omega \to \pi^+\pi^-\gamma$).

Измерение спиновых параметров в упругом рассеянии и бинарных реакциях рождения странных частиц впервые позволит осуществить однозначный парциально-волновой анализ возбужденных состояний нуклона во всей резонансной области и, тем самым, восстановить истинный спектр и характеристики барионных резонансов и, также, константы связи с различными каналами распада, то есть получить надежную базовую экспериментальную информацию для развития теоретических представлений о силах взаимодействия между конституэнтами нуклона. Измерения спиновых характеристик реакций двухчастичного и квазидвухчастичного рождения странных частиц позволят осуществить безмодельный амплитудный анализ этих процессов и принесут важную информацию о механизмах взаимодействия в странном секторе и, возможно, объяснения некоторых не понятых до настоящего времени экспериментальных фактов.

Во второй части программы представлена группа предложений, направленных на изучение влияния ядерной среды на процессы взаимодействия адронов, характеристики резонансов, а также проявлений ненуклонных степеней свободы и флуктуаций плотности в ядрах.

Адекватное описание поведения адронов в ядерной среде остается весьма актуальным для физики частиц. В настоящее время достаточно хорошо теоретически и экспериментально изучена роль квазиупругих процессов при прохождении адронов через ядро. Однако, этого нельзя сказать о неупругих взаимодействиях; недавний анализ первых данных по двойной перезарядке пиона на ядрах при ГэВ-ных энергиях показал, что этот процесс предоставляет уникальную возможность наблюдать эффект неупругих глауберовских перерассеяний как основной, уже начиная с $\approx 0.7$ ГэВ. Действительно, в области этих энергий обычный двухступенчатый механизм этой реакции с однопионным промежуточным состоянием не доминирует из-за малости сечения однократной перезарядки на нуклоне. Таким образом, впервые становится возможным в явном виде исследовать неупругие механизмы многократного взаимодействия адронов с нуклонами при их прохождении через ядро.

Предполагаемое явление ядерной критической опалесценции тесно связано с пионными степенями свободы и объясняет наблюдения узкого пика погло-щения пионов с энергиями вблизи 50 МэВ в ядрах. Другая возможность интерпретации этого пика — существование $d'$-дибариона.

Пространственно-временная картина глубоконеупругих процессов взаимодействия адронов с ядрами может быть восстановлена интерференционными методами и содержит информацию о продольных и поперечных размерах области взаимодействия, флуктуациях ядерной плотности и возможных проявлениях кварковых мешков.

Исследования характеристик подпорогового рождения частиц, определяемых свойствами ядерной материи на малых межнуклонных расстояниях, позволят получить информацию о спектральных функциях ядер в области, недоступной для изучения на электронных ускорителях. Сильная зависимость сечений подпороговых реакций от массы рожденной системы делает их чувствительным инструментом для поиска предсказанных теорией эффектов модификации свойств адронов в ядерной среде.

Рождение и распространение мезонных резонансов в ядрах позволит получить сведения о характеристиках резонансов в ядерной среде и, в результате изучения взаимодействий в конечном состоянии, также о процессах рассеяния мезонных резонансов на нуклонах и взаимодействий между $\Lambda$-гиперонами и $\Delta$-изобарами.

Следует отметить, что выполнение представленной программы измерений на синхротроне ИТЭФ привело бы также к получению экспериментальных данных, необходимых для однозначной интерпретации результатов, ожидаемых на новых ускорителях CEBAF и RHIC. Например, исследования возбуждения нуклонов адронами (ИТЭФ) и фотонами (CEBAF) дополняют друг друга и позволяют извлечь относительные вероятности радиационных распадов резонансов, а для надежной идентификации сигналов кварк-глюонной плазмы нужны подробные данные о рождении различных адронов в адрон-ядерных и ядро-ядерных процессах.

Ближайшее будущее адронной физики в мире, несмотря на ее чрезвычайно важное значение для развития теории сильных взаимодействий на расстояниях порядка размера нуклона, представляется очень неопределенным. Планируемый в КЕК (Япония) новый протонный ускоритель JHF (Japanese Hadron Facility), перекрывающий область промежуточных энергий, не будет введен в строй по крайней мере в текущем десятилетии. **За это время в ИТЭФ можно было бы в условиях слабой конкуренции получить ценную информацию по принципиальным аспектам адронной физики.**

**Предлагаемая экспериментальная программа** частично базируется на существующих экспериментальных установках, но в своей основной части **требует создания нового магнитного спектрометра, оснащенного быстрыми координатными и гамма-детекторами, а также поляризованными мишенями.**

Таблица 3 подытоживает существующие возможности и потребности предложенных экспериментов в различных ресурсах, рассматриваемых в

Таблица 3: Сводная таблица предлагаемых экспериментов

| Эксперимент | Режим ускорителя | | Наличие установки | Реализация в рамках общего спектрометра | | |
|---|---|---|---|---|---|---|
| | Высокая площадка | Медленный вывод ионов | | На I этапе | На II этапе | В перспективе |
| 2.1. Поиски и исследование экзотических состояний | √[2] | | | √ | √ | √ |
| 2.2. Исследование низколежащих скалярных мезонов | √ | | | | √ | |
| 2.3. Измерение параметров вращения спина | √ | | √ | | | |
| 2.4. Исследование структуры $\eta$- и $f_0$-мезонов [1] | √ | | | | | |
| 2.5. Проверка правила OZI | √ | | | | √ | |
| 2.6. Поляризационные параметры в реакциях рождения странных частиц | √ | | | | √ | |
| 2.7. Поиск аномальных явлений в пороговом рождении пар гиперонов [1] | √ | | | | | |
| 2.8. Поиск распада $\omega \to \pi^+\pi^-\gamma$ | √ | | | | √ | |
| 2.9. Измерение электромагнитных формфакторов $\eta$-, $\eta'$- и $\omega$-мезонов | √ | | | | | √ |
| 3.1. Оценка размеров флуктона | √ | | | √ | | |
| 3.2. Исследование взаимодействия флуктонов | √ | √ | | √ | | |
| 3.3. Исследование свойств ядерной материи в подпороговом образовании адронов [1] | | | √ | | | |
| 3.4. Исследование явления ядерной критической опалесценции | [3] | | | √ | | |
| 3.5. Исследование "выброса" ядерной материи | √ | √ | | | | |
| 3.6. Двойная перезарядка пионов на ядрах | √ | | | √ | | |
| 3.7. Изучение характеристик векторных и скалярных мезонов | √ | | | | √ | √ |
| 3.8. Исследование адрон-ядерных взаимодействий в области, далекой от квазисвободной кинематики | √ | | | √ | | |
| 3.9. Исследование структуры ядер | √ | √ | | | | |

[1] Эти эксперименты предполагается проводить на внутреннем пучке ускорителя.
[2] Кроме 2.1..5.
[3] На втором этапе необходимы пионные пучки.

рамках настоящей программы.

Ввиду весьма серьезных трудностей с финансированием науки в России, очень важно, что **большинство предложений может быть реализовано на одной общей экспериментальной установке,** что позволит сконцентрировать имеющиеся средства и людские резервы. Это соответствует рекомендации Комиссии ИТЭФ по экспертизе экспериментов на ускорителе ИТЭФ (К.Г. Боресков, В.Б. Гаврилов, А.И. Голутвин и А.Б. Кайдалов). **Институт располагает существенной частью необходимой материальной базы** (в частности, за основу спектрометра общего пользования может быть взят модернизированный магнит имеющегося трехметрового спектрометра, а также использована криовакуумная инфраструктура для работы поляризованных мишеней), **и большим опытом создания аналогичных установок. Ввод в действие этой установки может быть осуществлен поэтапно.** По мере комплектования спектрометра детекторами различных типов **может быть реализована часть экспериментов,** не требующих полной комплектации.

Не менее важным является вопрос о восстановлении режима работы ускорителя У-10 ИТЭФ на полной проектной энергии 10 ГэВ, необходимой, в частности, для получения качественных пионных пучков. **Большинство предложенных экспериментальных программ подразумевает работу ускорителя на высокой энергетической площадке.**

Программа исследований на ускорителе ИТЭФ составлена из предложений, представленных лидерами и ведущими сотрудниками экспериментальных групп (см. нижеследующую таблицу).

| Раздел | Фамилия И.О. |
|---|---|
| Введение | В.В. Владимирский |
| 2.1.1.–3., 2.1.6., 2.5., 2.9., 3.7. | В.Т. Смолянкин |
| 2.1.4. | В.Л. Столин |
| 2.1.5. | Л.С. Воробьев, Г.А. Лексин |
| 2.2. | В.П. Канавец |
| 2.3. | В.П. Канавец, В.В. Сумачев (ПИЯФ) |
| 2.4. | В.П. Чернышев, А.Г. Долголенко |
| 2.6. | В.В. Рыльцов |
| 2.7. | В.С. Борисов, А.Г. Долголенко |
| 2.8. | В.В. Куликов |
| 3.1., 3.2. | Г.А. Лексин |
| 3.3. | Ю.Т. Киселев, В.А. Шейнкман |
| 3.4. | М.В. Косов |
| 3.5. | В.А. Смирнитский |
| 3.6. | А.П. Крутенкова, В.В. Куликов |
| 3.8. | Н.А. Пивнюк, В.М. Колыбасов (ФИАН) |
| 3.9. | Ю.В. Требуховский |
| 4., табл. 3 | И.Г. Алексеев, Д.Н. Свирида |
| Заключение | В.П. Канавец |

Предложения, составляющие настоящую программу, открыты для присоединения как отечественных, так и иностранных участников.